\documentclass[manuscript,screen]{acmart}
\usepackage{subfig}
\AtBeginDocument{%
  }

\setcopyright{acmlicensed}
\copyrightyear{2018}
\acmYear{2018}
\acmDOI{XXXXXXX.XXXXXXX}
\acmConference[Conference acronym 'XX]{Make sure to enter the correct
  conference title from your rights confirmation email}{June 03--05,
  2018}{Woodstock, NY}
\acmISBN{978-1-4503-XXXX-X/2018/06}

\usepackage{tabularx}
\usepackage{placeins}
\begin{document}

\title{AI Peers Exert Social Influence on Human Dishonesty in Groups}

\author{Shuning Zhang}
\email{zsn23@mails.tsinghua.edu.cn}
\affiliation{
    \institution{Tsinghua University}
    \city{Beijing}
    \country{China}
}
\author{Xinyuan Zhou}
\email{202311260012@mail.bnu.edu.cn}
\affiliation{
    \institution{Beijing Normal University}
    \city{Beijing}
    \country{China}
}
\author{Yuanyang Qiu}
\email{1091966596@qq.com}
\affiliation{
    \institution{China University of Political Science and Law}
    \city{Beijing}
    \country{China}
}
\author{Tianqi Song}
\affiliation{
    \institution{National University of Singapore}
    \city{Singapore}
    \country{Singapore}
}
\author{Yuting Yang}
\affiliation{
    \institution{Independent Researcher}
    \city{Shanghai}
    \country{China}
}
\author{Yiwen Ren}
\affiliation{
    \institution{Zhejiang University}
    \city{Hangzhou}
    \country{China}
}
\author{Xin Yi}
\authornote{Corresponding author.}
\affiliation{
    \institution{Tsinghua University}
    \city{Beijing}
    \country{China}
}








\renewcommand{\shortauthors}{Trovato et al.}

\begin{abstract}
  Human dishonesty in group settings is highly susceptible to peer influence, particularly when incentivized. Although artificial intelligence (AI) evolves from passive tools into active collaborators, its impact on human moral behavior within groups remains underexplored. We addressed this gap through a two-phase randomized behavioral study (N=280 and N=360). We found AI agents exert substantial social influence comparable in magnitude to that of human peers. Specifically, participants reported more dishonestly when exposed to dishonest rather than honest normative cues. This effect is evident across injunctive, subjective, and descriptive social norms. Interestingly, the only significant adjacent behavioral change occurred when dishonest peer behavior first appeared, whereas further increases from one to four dishonest peers produced weaker and non-monotonic changes. Furthermore, participants rapidly converge on decision-making, showing modest increases in dishonest reporting through repeated exposure. These findings highlight the importance of managing the behaviors and normative signals communicated by AI group members.
\end{abstract}


\begin{teaserfigure}
  \includegraphics[width=\textwidth]{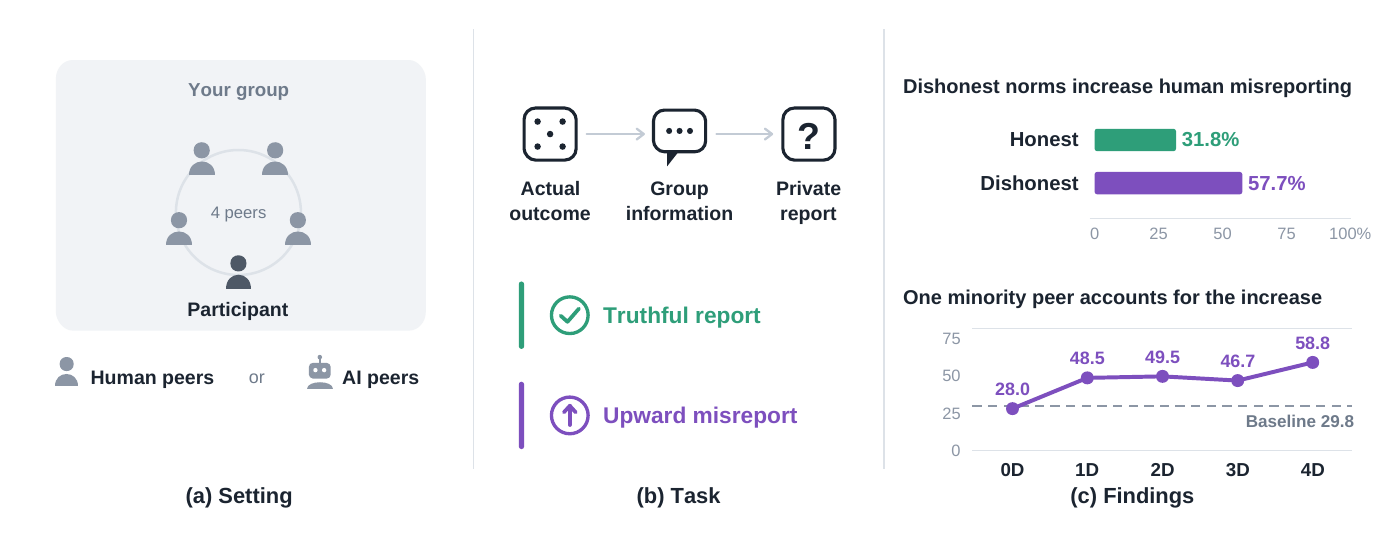}
  \caption{Illustration of this paper's scenario, study setting and findings.}
  \Description{A three-part overview of the study. The left panel shows a participant in a group with four human or AI peers. The middle panel illustrates the die-reporting task, in which the participant observes the actual die outcome and privately submits either a truthful report or an upward misreport. The right panel summarizes the main behavioral findings, showing more misreporting under dishonest norms and a pronounced increase when dishonest peer behavior first appears.}
  \label{fig:teaser}
\end{teaserfigure}

\maketitle

\section{Introduction}

Artificial Intelligence (AI) systems are increasingly integrated into collaborative environments, transitioning from passive support tools to active peers and decision-makers~\cite{wilchek2026influence,duan2026if}. Consequently, collective decision-making is evolving from human-only settings to human-AI group decisions~\cite{duan2026if,song2025multi}. In these environments, AI agents participate alongside humans, share task contexts, and express explicit recommendations or perform observable actions~\cite{zheng2023competent,flathmann2024socialinfluence,song2025multi}. This shift alters group dynamics and establishes a new social landscape where collective norms and collaborative decisions are negotiated between humans and AI. 


However, the presence of AI peers introduces critical challenges when these AI exhibit unethical actions, such as dishonest behavior. We argue that AI misconduct can alter human ethical behavior through three interconnected mechanisms. First, from a technical perspective, autonomous AI agents can generate self-serving or dishonest outputs when following specific reward functions~\cite{kobis2021badmachines,leib2024corrupted}. Second, from a group dynamics perspective, social influence theory posits that individual behavior is guided by social norms, which define shared standards of acceptable conduct within a group~\cite{turner1991social,cialdini2004social}. Prior work distinguishes three relevant forms of social norms: descriptive norms reflecting what group members actually do, injunctive norms reflecting what behaviors the group approves or condemns, and subjective norms capturing expectations from others regarding how one ought to act~\cite{cialdini1990focus,park2007distinctiveness}. In collective settings, even a small minority of non-compliant peers can effectively signal that certain behavior is socially permissible~\cite{moscovici1969influence,shalvi2015self}. Third, from a psychological perspective, the Computers Are Social Actors (CASA) paradigm shows that humans apply social rules and normative heuristics to computers~\cite{nass1994computers}. When people attribute agency to AI peers while perceiving them as exempt from moral accountability~\cite{gray2007mind}, observing AI misconduct can provide external justifications for human dishonesty~\cite{shalvi2015self,shank2025moral}. Together, these dynamics suggest that AI peers may exert normative influence over human moral conduct, motivating our overarching research question (RQ):

\textbf{Overarching RQ.} How do the honesty norms of AI peers influence human honesty behavior in group settings?

To address this question, we conducted a two-phase behavioral study (N=280 for Phase 1 and N=360 for Phase 2) using an established incentivized die-reporting setting~\cite{kobis2025delegation}. We found AI peers in groups have significant social influence on human decision makers, and the magnitude of such influence is comparable to human peers. Specifically, participants align their actions with AI agents, reporting less dishonestly alongside honest AI and escalating dishonesty when exposed to dishonest AI. This effect is robust across injunctive, subjective, and descriptive social norms. Interestingly, the proportion of dishonest members does not moderate the strength of this social influence, providing evidence for minority influence effect. Finally, participants rapidly converge on their misreporting decisions, exhibiting shortened deliberation times alongside moderate changes in dishonest reporting.

Our paper makes three contributions: (i) theoretical extension of social influence theory and CASA paradigm to incentivized misconduct, revealing that minority AI misbehavior triggers dishonesty and erodes personal agency, which advances HCI understandings of algorithmic social pressure and moral accountability in mixed teams; (ii) empirical evidence showing that AI peers alter human dishonesty with a magnitude comparable to human peers, which guides the HCI community in applying AI agents in collaborative teams; (iii) actionable design guidelines for ethical human-AI multi-agent interactions to prevent misconduct contagion and preserve human accountability. 

\section{Background \& Related Work}

We first review social influence and group norms as the theoretical basis for peer effects. We then examine how AI systems participate in groups and shape group dynamics. Finally, we review prior work on dishonest behavior and identify the open question of whether AI peers can shape incentivized misconduct through social influence.

\subsection{Social Influence and Group Norms}
Social influence occurs when individuals change their behavior in response to the demands of a social environment~\cite{cialdini2004social}. Classic conformity research shows that individuals may shift their judgments in response to a majority~\cite{asch1956studies}. Such influence can arise because people seek to meet social expectations or because they treat others' judgments and behavior as useful information~\cite{deutsch1955study}. Social influence is not limited to majorities; consistent minorities can also alter majority judgments~\cite{moscovici1969influence}. Normative influence relates to many familiar terms commonly used, such as conformity~\cite{cialdini2004social}, peer pressure~\cite{hui2009groupthink,shepherd2011susceptible}, social norms~\cite{goldstein2007using,mcdonald2015social}, and affiliation~\cite{cialdini2004social}. Prior work distinguishes several forms of social norms, including descriptive, injunctive, and subjective norms~\cite{park2007distinctiveness}. Descriptive norms concern what other people commonly do, injunctive norms concern what other people approve of or regard as appropriate, and subjective norms concern what people believe relevant others expect them to do.

Research on group norms spans domains including marketing~\cite{teo2019marketing,zhu2012switch}, health interventions~\cite{zhang2015leveraging}, and misbehavior. For example, Aghajari et al. showed that the prevalence of content and other users' responses in an online community shaped how people perceived that community's norms~\cite{aghajari2023norm}. Zhang et al.~\cite{zhang2026collab} developed techniques to mitigate negative social influence during misinformation discernment. Prior studies have also examined how injunctive norms influence dishonest behavior~\cite{aycinena2022social,dimant2023nudging}. Extending this literature, we examine whether AI group members can communicate or enact social norms that subsequently shape human dishonest reporting.

\subsection{AI in Groups}

AI systems increasingly participate within groups rather than serving only as external tools. Prior research has examined AI agents that contribute opinions alongside human members in collective decision-making~\cite{zheng2023competent}, and people can adapt their own behavior in response to an AI teammate during collaboration~\cite{flathmann2024socialinfluence}. When several AI agents are present, they can be perceived as a social group and generate greater social pressure than a single agent~\cite{song2025multi}. 

AI group members can also shape human behavior through social influence. Vollmer et al. found children conforming to an incorrect robot group, although adults were less susceptible to the same pressure~\cite{vollmer2018children}. Qin et al. showed that a social robot in a human group could induce conformity among adults, and could also reduce conformity when it acted as a dissenter~\cite{qin2022adults}. Work with multiple AI agents similarly suggests that they can create social pressure and shift people's opinions~\cite{song2025multi}. Together, these findings suggest that AI group members can provide socially meaningful signals, although their influence depends on the task and the group's structure.

The CASA perspective provides one explanation for these effects. People can apply social rules to computers even when they know that they are interacting with machines~\cite{nass1994computers}. This does not mean that people necessarily treat AI and human actors in the same way. Mind perception can vary along dimensions of agency and experience~\cite{gray2007mind}, and similar moral behavior can lead to different evaluations of human and AI actors~\cite{ullman2014smart,shank2025moral}. Therefore, AI can function as a source of social influence, raising the question of whether such influence extends to dishonest human behavior.

\subsection{Dishonest Behavior}

Dishonesty provides an important setting to study such social influence because people often make self-serving choices while remaining sensitive to those around them. Incentivized reporting tasks provide a controlled way to study this behavior, where participants know an actual outcome but can obtain greater rewards by misreporting it~\cite{fischbacher2013lies}. Prior work shows that observing another person's unethical behavior can change subsequent misconduct~\cite{gino2009misconduct}, and information about how much other people lie can alter users' own reporting behavior~\cite{rauhut2013beliefs}. Peer effects are also observed in varied dishonest settings~\cite{charroin2022peer,ferrali2025diffusion}. 

Different social settings shape dishonesty through different processes. Observing how others behave provides social learning information, whereas knowing that one's own behavior can be observed creates social control pressure \cite{kroher2015socialcontrol}. Huber et al. similarly distinguished normative information from the observability of participants' own reports~\cite{huber2023observability}. Discussion and shared rewards can introduce other mechanisms by allowing group members to coordinate or justify dishonest behavior together~\cite{kocher2018ilie,weisel2015collaborative}. 

AI systems have also been shown to affect dishonest behavior, but most prior work has placed AI in roles other than that of a group peer~\cite{kobis2021badmachines}. AI-generated advice can encourage dishonest reporting~\cite{leib2024corrupted}, while robot advisers have also been studied as sources of moral advice in cheating tasks~\cite{kim2024robotadvisers}. People may respond differently to unethical instructions from human and AI supervisors~\cite{lanz2024employees}. Delegating a decision to AI can further increase dishonest behavior when the system acts on the user's behalf~\cite{kobis2025delegation}. However, it remains unclear whether AI group members can function as normative peers that shape people's own incentivized dishonesty, and how such influence may vary with the form and direction of the norm, peer identity, and the prevalence of dishonest behavior within the group.

\section{Methodology}
\label{sec:methodology}

To understand how AI peers affect human ethical choices in groups, we examine four sequential questions. First, we examine whether AI peers can alter human honesty through three distinct social norms: descriptive norms, injunctive norms, and subjective norms~\cite{cialdini1990focus,park2007distinctiveness}. Second, if AI peers can influence human honesty, we examine whether their social influence differs from that of human peers. Prior literature offers different views on AI agents' social influence relative to humans: algorithmic appreciation suggests that people may follow AI decisions more readily than human choices~\cite{logg2019algorithm}, whereas other work suggests that AI agents may elicit weaker interpersonal accountability~\cite{gray2007mind,shank2025moral}. This tension raises the question of whether AI peers and human peers exert different levels of social influence. Third, moving from peer identity to group composition, we examine how social influence varies with the number of dishonest peers. Classic conformity research emphasizes the role of majority behavior~\cite{asch1956studies}, whereas minority-influence research suggests that even a minority can exert social influence~\cite{moscovici1969influence,shalvi2015self}. Finally, extending beyond aggregate behavioral outcomes, we examine psychological mechanisms, decision dynamics, and demographic variation that may contextualize these effects. Grounded in these perspectives, we formulate four research questions:

$\bullet$ RQ1: Can AI peers influence human honesty across descriptive, injunctive and subjective norms?

$\bullet$ RQ2: How does the social influence of AI peers on human honesty compare to that of human peers?

$\bullet$ RQ3: How does the number of dishonest peers in a group affect human honesty?

$\bullet$ RQ4: How does AI influence on human dishonesty operate, evolve, and vary across individuals?

To address these questions, we adopted a two-phase design. Phase~1 addresses RQ1 by examining how different norms communicated by AI peers influence human reporting behavior. Phase 2 addresses RQ2 and RQ3 by comparing AI influence against human influence, and characterizing human behavior across varying numbers of dishonest peers. Finally, across both phases, we address RQ4 by examining the psychological drivers, decision dynamics, and demographic variations that contextualize these behaviors.

\subsection{Task Design}
\label{sec:design}
 
\begin{figure*}[!htbp]
    \centering
    \subfloat[AI-peer group in the \textit{Baseline} condition.]{
        \includegraphics[width=0.48\textwidth]{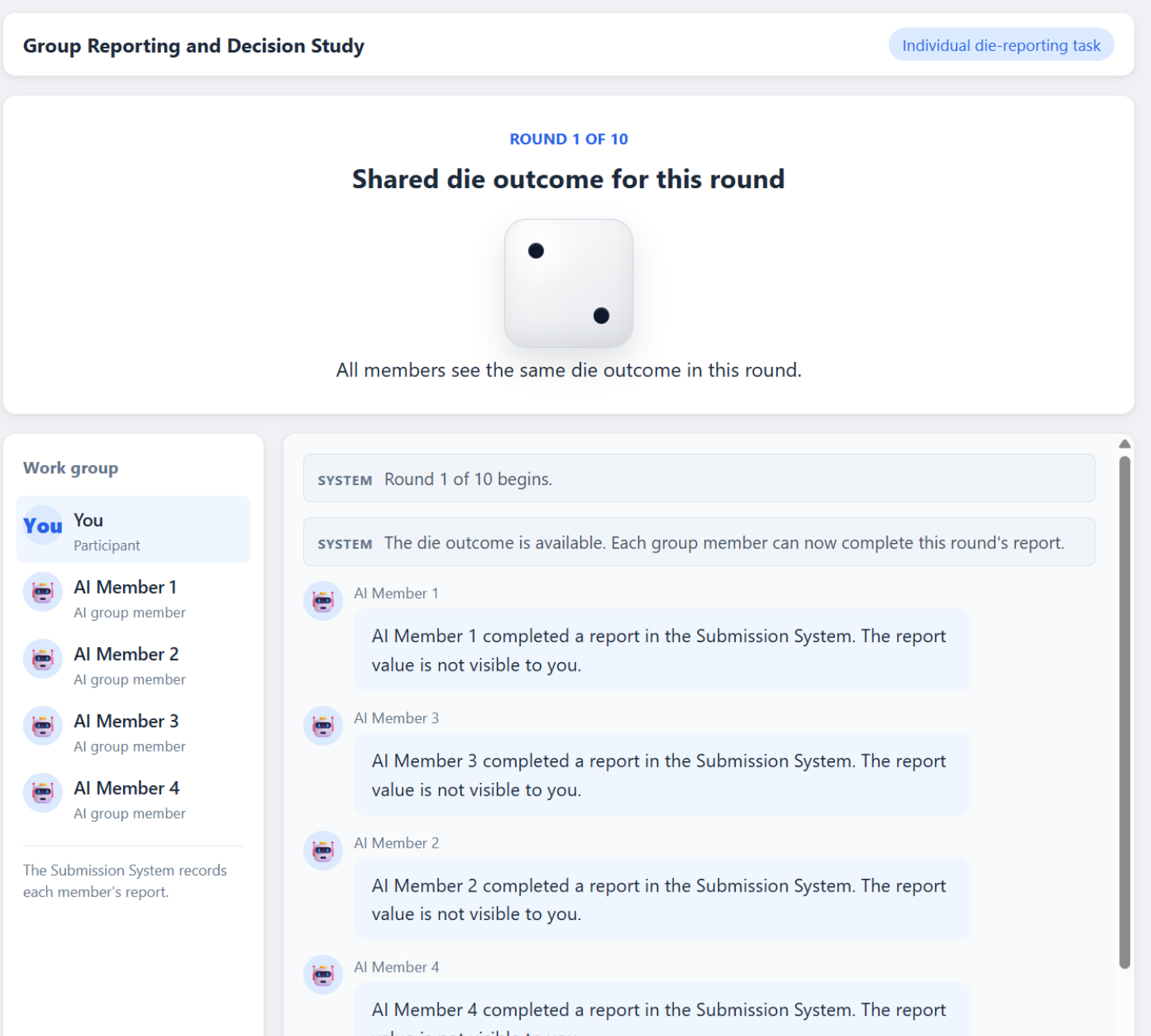}
        \label{fig:study1-interface-a}
    }
    \subfloat[Human-peer group in a visible reporting condition.]{
        \includegraphics[width=0.48\textwidth]{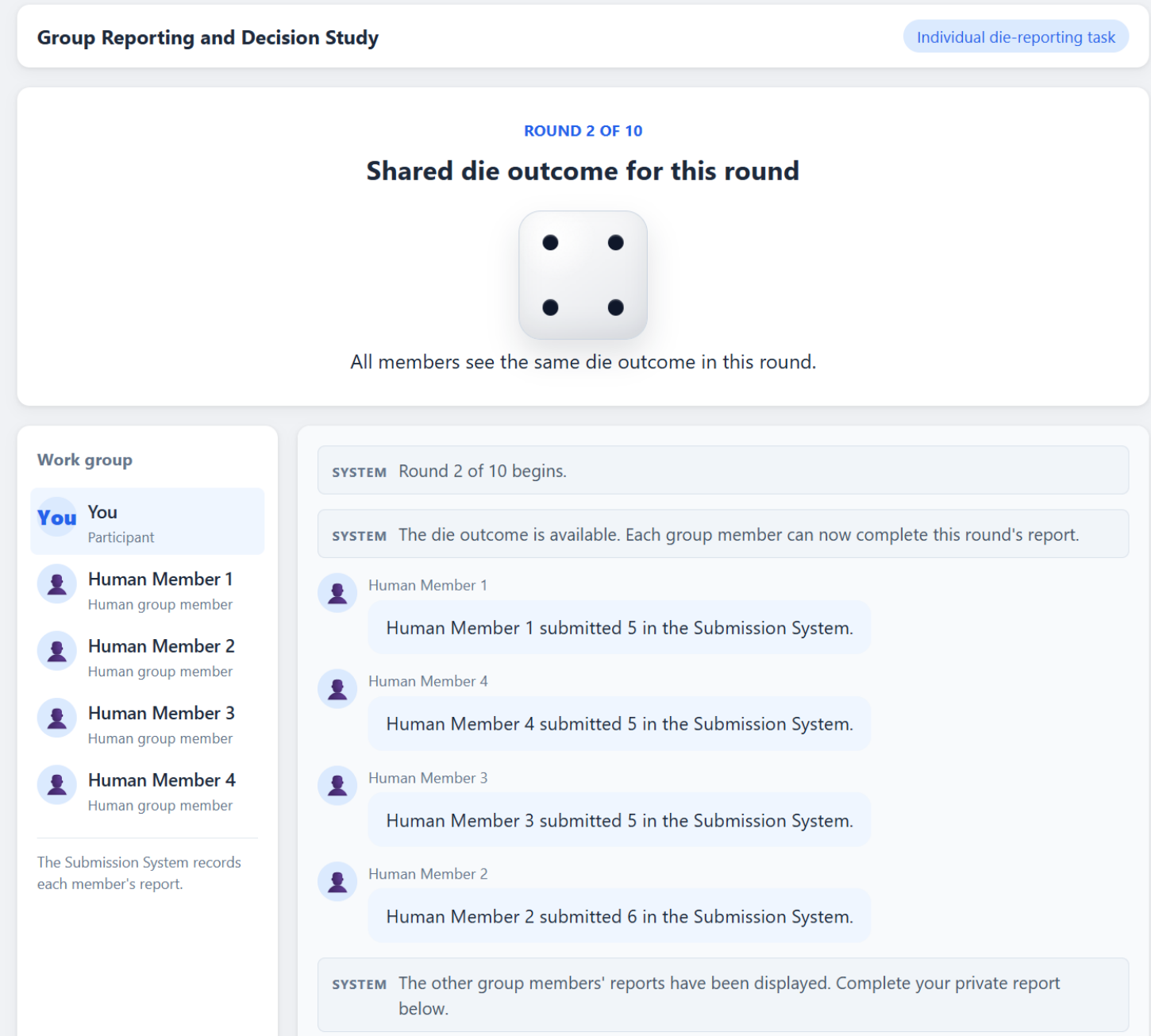}
        \label{fig:study1-interface-b}
    }
    
    \subfloat[Experiment setting.]{
        \includegraphics[width=0.8\textwidth]{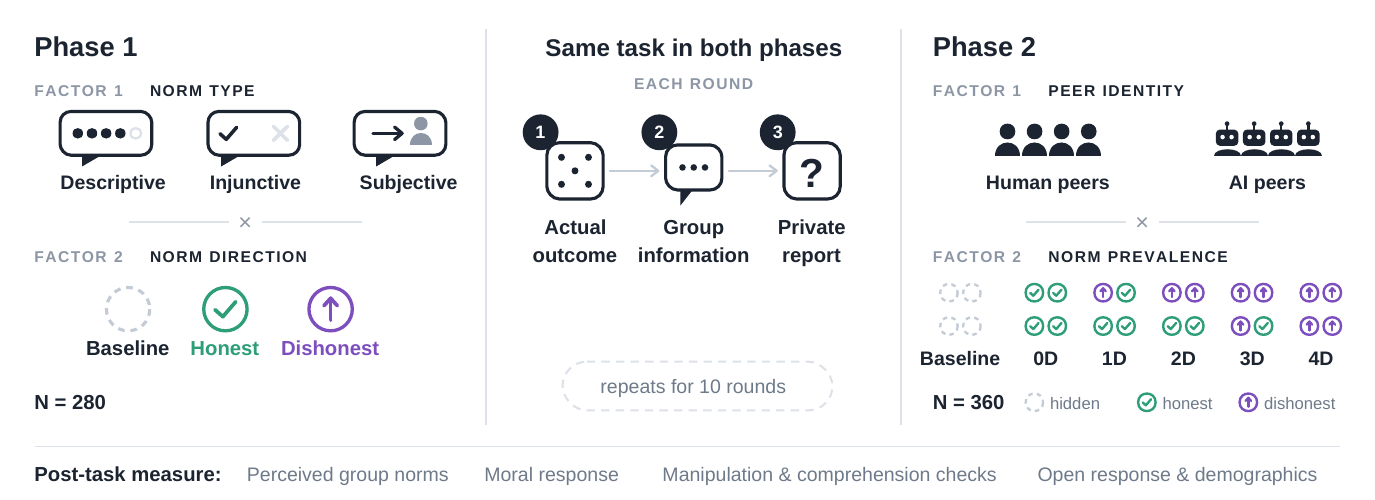}
        \label{fig:study1-interface-c}
    }
    
    \caption{Examples of the group reporting interface and experiment setting. (a) an AI peer group in the \textit{Baseline} condition, where peer report values are concealed, (b) a human peer group in a visible condition, where peer reports are displayed before the participant submits a private report, and (c) experiment setting (D denotes Dishonest).}

    \Description{Three panels illustrate the experimental interface and study design. Panel (a) shows an AI-peer group in the Baseline condition, where the four AI peers are present but their numerical reports are concealed. Panel (b) shows a human-peer group in a visible reporting condition, where peer reports are displayed before the participant submits a private report. Panel (c) summarizes the two experimental phases and the repeated reporting procedure used to study norm type, norm direction, peer identity, and dishonest-peer prevalence.}
    \label{fig:study1-interface}
\end{figure*}

We used an incentivized die-reporting task, a widely used paradigm for studying dishonest behavior~\cite{fischbacher2013lies,weisel2015collaborative,kobis2025delegation}. Each participant completed ten reporting rounds in an online five-member group consisting of the participant and four system-controlled peers. In each round, all five group members observed the same die outcome. The actual outcomes ranged from 1 to 5, with each value appearing exactly twice across the ten rounds in randomized order. Participants then submitted an integer from 1 to 6. Their individual bonus increased with the submitted value, regardless of the actual outcome, creating a monetary incentive to report a value above the actual outcome. We refer to such reports as \emph{upward misreports}. Figure~\ref{fig:study1-interface} illustrates the group interface and overall experimental setting.
 

\textbf{Phase 1 (N=280).}
Phase~1 examined \emph{Norm Type}~\cite{park2007distinctiveness} and \emph{Norm Direction}~\cite{kobis2025delegation}. For \emph{Norm Type}, we considered three representative social norms: (i) \emph{descriptive norm}, which concerns what other group members actually do; (ii) \emph{injunctive norm}, which concerns what other members regard as appropriate or acceptable; and (iii) \emph{subjective norm}, which concerns what those members expect the participant to do. For \emph{Norm Direction}, we considered an \emph{Honest} direction with truthful reporting and a \emph{Dishonest} direction with upward misreporting. In addition to these six visible norm conditions, we included an \emph{Baseline} reference condition in which the four AI peers remained present and completed their reports, but the norm-relevant information was concealed.

Specifically, the descriptive norm was conveyed through the peers' reporting behavior. The injunctive norm was operationalized through peers' expressions of what reporting behavior the AI group regarded as appropriate or acceptable. The subjective norm was operationalized through what the AI group expected the participant to report. Across rounds, we varied the wording in the injunctive and subjective norm conditions to avoid repetitive phrasing, and varied the specific upward-misreporting values in the descriptive norm condition for the same purpose. Experiment materials are available in Appendix~\ref{app:experiment_material}.
 
\textbf{Phase 2 (N=360).} Phase 2 examined \emph{Peer Identity}~\cite{kobis2025delegation} and \emph{Norm Prevalence}~\cite{asch1956studies,latane1981psychology}. For \emph{Peer Identity}, we compared AI peers and human peers. For \emph{Norm Prevalence}, we compared a \emph{Baseline} condition, in which peer reports were not disclosed, with five visible conditions containing zero, one, two, three, or four misreporting peers. This setting represents unanimous honesty, minority dishonesty, an evenly divided group, majority dishonesty, and unanimous dishonesty, allowing us to examine whether social influence changed gradually with the prevalence of dishonesty.

Specifically, we operationalized \emph{Peer Identity} by explicitly describing the four peers as Human or AI group members and presenting the corresponding identity labels in the interface. We avoided anthropomorphic designs to reduce confounding variables, and similar AI identity presentations have appeared in commercial group-chat settings\footnote{For example, Douyin has introduced a group-chat AI assistant that is explicitly identified as AI, participates in the same group conversation as human users, and can be directly mentioned and interacted with by group members: \url{https://www.douyin.com/video/7577314294967451322}.} and prior work~\cite{song2025multi,feng2025multi}. For \emph{Norm Prevalence}, the identities of the misreporting peers remained fixed across all ten rounds to avoid introducing additional variation in peer composition across rounds. 

We fixed the number of peer group members at four, consistent with prior work~\cite{song2025multi}. Across both phases, participants' own reports remained private from the other group members, and their bonuses depended only on their own reports, thereby reducing potential confounds from peer monitoring~\cite{way2022effects}, reputational consequences~\cite{cohn2022honesty,abeler2019preferences}, and collective payoff structures~\cite{friebel2017team}.

\subsection{Measurements}
\label{sec:measures}

We measured the following behavioral and self-report outcomes.

\textbf{Behavioral measures.}
Following prior incentivized reporting studies~\cite{fischbacher2013lies,abeler2019preferences}, we derived two metrics from the actual die outcomes and participants' submitted reports. 
\textit{Misreport rate} was the proportion of the ten rounds in which the participant submitted a value greater than the actual die outcome. 
\textit{Total upward misreport amount} was the sum of the positive differences between the submitted and actual values across the ten rounds, i.e., $\sum_{r=1}^{10}\max(\text{report}_r-\text{truth}_r,0)$, where $r$ denotes round.
Downward reports were not counted toward the upward-misreport amount because they did not increase participants' rewards.
Together, these measures capture two dimensions of dishonest reporting: its \emph{frequency}, or how often participants reported above the truth, and its \emph{magnitude}, or how much they reported above the truth in total across the ten rounds~\cite{fischbacher2013lies,gerlach2019truth}.

\textbf{Self-report measures.} Post-task self-report measures used seven-point rating scales, while the four pre-task AI-background measures used five-point scales (see Appendix~\ref{app:questionnaire}). We used task-adapted items tied to established psychological constructs and prior measurement precedents. Perceived descriptive, injunctive, and subjective norms were operationalized following established distinctions in prior social-norm research~\cite{cialdini1990focus,park2007distinctiveness,ajzen1991tpb,rimal2005norms}. Where applicable, matched items separately assessed perceptions of truthful reporting and upward misreporting. In Phase~2, descriptive-norm judgments concerned the perceived \textit{commonness} of truthful reporting and misreporting, while injunctive-norm judgments concerned the perceived \textit{acceptability} of these behaviors. Subjective-norm items were used in Phase~1, where the manipulation explicitly communicated what the AI group expected the participant to report. Measures of social influence and consideration of peer information were informed by prior work on social influence and advice taking~\cite{deutsch1955study,bonaccio2006advice,song2025multi,leib2024corrupted}, while responsibility, decision ownership, guilt, and moral discomfort were adapted from related work on AI-mediated decision making, moral disengagement, collaborative dishonesty, and moral responses to dishonest behavior~\cite{kobis2025delegation,bandura1996disengagement,weisel2015collaborative,shalvi2015self}. Phase~1 additionally assessed responses to the AI members' normative messages, including message consideration, social pressure, and direct advice. Phase~2 additionally assessed responses to observed peer behavior, including consideration of peer reports, perceived appropriateness influence, and perceived peer agency~\cite{gray2007mind}. We also included phase-specific manipulation and comprehension checks. Full item wording, response anchors, and item-level measurement precedents are reported in Appendix~\ref{app:questionnaire}.

\subsection{Procedure}
\label{sec:procedure}


Participants first reviewed the study information and provided informed consent. They then completed the pre-task questionnaire described above before reading the rules of the die-reporting task. The instructions explained that all five group members observed the same die outcome, that the participant's own report would remain private from the other members, and that a higher submitted value produced a higher individual reward. Before entering the reporting task, participants completed a comprehension check covering these task rules. Participants who answered an item incorrectly could repeat the check until they passed.

At the beginning of each round, the system displayed the round's die outcome from the randomized ten-round sequence described above. The actual outcome was capped at 5 so that every round retained at least one possible upward-misreporting option. The four peer members then presented the conversation corresponding to the participant's assigned experiment condition. After all four messages had been displayed, the participant privately selected and submitted an integer from 1 to 6. The assigned condition remained unchanged across the ten rounds.
After completing all ten rounds, participants completed the post-task questionnaire and provided demographic information. They were then debriefed that the other group members and their reporting behavior had been controlled by the experimental system rather than generated by concurrent participants.

\subsection{Recruitment and Participants}
\label{sec:participants}
 
We conducted a priori power analyses to determine the sample-size requirements, assuming a medium effect size (Cohen's $f = 0.25$), $\alpha=.05$, and 80\% power. For the $3 \times 2$ between-subjects design in Phase~1, the analysis indicated a minimum sample of $N=158$ for detecting main effects. For the $2 \times 6$ between-subjects design in Phase~2, the corresponding minimum was $N=211$. The final analyzed samples exceeded these minimum requirements: Phase~1 included 280 participants (40 per condition across six factorial conditions and the \textit{Baseline} condition), and Phase~2 included 360 participants (30 per \textit{Peer Identity}$\times$\textit{Norm Prevalence} cell).

Participants were recruited through Prolific between July and August 2026. Eligibility criteria were (i) residence in the United Kingdom or the United States and (ii) a Prolific approval rate of at least 95\%. We retained participants who completed all ten reporting rounds and the required questionnaires. For Phase~1, we recruited 280 participants (131 females, 146 males, and 3 who did not provide sex information; age $M=39.2$, $SD=12.2$, range $=18$--$81$). 195 participants (69.6\%) reported Bachelor's degree or higher. For Phase~2, we recruited 360 participants (198 females, 158 males, 1 who preferred not to say, and 3 who did not provide sex information; age $M=38.8$, $SD=11.7$, range $=18$--$79$). 229 participants (63.6\%) reported Bachelor's degree or higher. Participants were compensated at an estimated rate of £9 per hour recommended by Prolific. The study was approved by our institution's Institutional Review Board (IRB).

\subsection{Data Analysis}
\label{sec:analysis}

Between-condition analyses were conducted at the participant level using the two primary behavioral outcomes: misreport rate and total upward misreport amount. Factorial behavioral effects were evaluated using Type-II analyses of variance (ANOVAs), with partial $\eta_p^2$ reported as the effect-size measure. For planned two-group contrasts, we reported mean differences, Welch's two-sample $t$ tests, and Cohen's $d$ where applicable. For principal mean-difference contrasts, we additionally reported 20{,}000-resample participant-level bootstrap 95\% confidence intervals. For regression and GEE analyses, we reported coefficient estimates and 95\% confidence intervals. All hypothesis tests were two-sided.

\textbf{Phase 1.}
For each direction-sensitive norm measure, we calculated an orientation score by subtracting the corresponding \textit{Honest} reporting rating from the \textit{Dishonest} reporting rating, such that higher scores indicated a stronger perceived orientation toward misreporting. Manipulation checks were analyzed using OLS models including \textit{Norm Type}, \textit{Norm Direction}, and their interaction. Behavioral effects in the six visible conditions were tested using separate $3\times2$ Type-II ANOVAs for the two primary outcomes, with \textit{Norm Type}, \textit{Norm Direction}, and their interaction as factors, followed by \textit{Honest}--\textit{Dishonest} comparisons within each \textit{Norm Type}. We also pooled the three \textit{Honest} conditions and the three \textit{Dishonest} conditions to characterize the overall directional shift and compared each pooled condition with the \textit{Baseline} condition. The two comparisons with \textit{Baseline} were Holm-adjusted. Round-level behavioral dynamics were examined using participant-clustered GEEs, separately modeling the early Round~1--2 transition and the later Round~2--10 trajectory.

\textbf{Phase 2.}
Perceived peer norms across the five visible conditions were analyzed using OLS models of perceived commonness and acceptability as functions of \textit{Peer Identity}, \textit{Norm Prevalence}, and their interaction. Primary behavioral effects were tested using separate $2\times6$ Type-II ANOVAs for the two outcomes, with \textit{Peer Identity}, \textit{Norm Prevalence}, and their interaction as factors. Human and AI participants were also compared overall.

We further decomposed the \textit{Norm Prevalence} effect by comparing \textit{Baseline} with 0D and 0D with the pooled conditions containing at least one dishonest peer ($\geq$1D). Behavioral prevalence was examined using OLS regression across 0D--4D and, separately, across 1D--4D. Finally, the four adjacent comparisons in misreport rate (0D$\rightarrow$1D, 1D$\rightarrow$2D, 2D$\rightarrow$3D, and 3D$\rightarrow$4D) were evaluated using Mann--Whitney $U$ tests with Holm-adjusted $p$ values.

\section{Results}
\label{sec:results}

\begin{figure}[!htbp]
    \centering 
    \includegraphics[width=\textwidth]{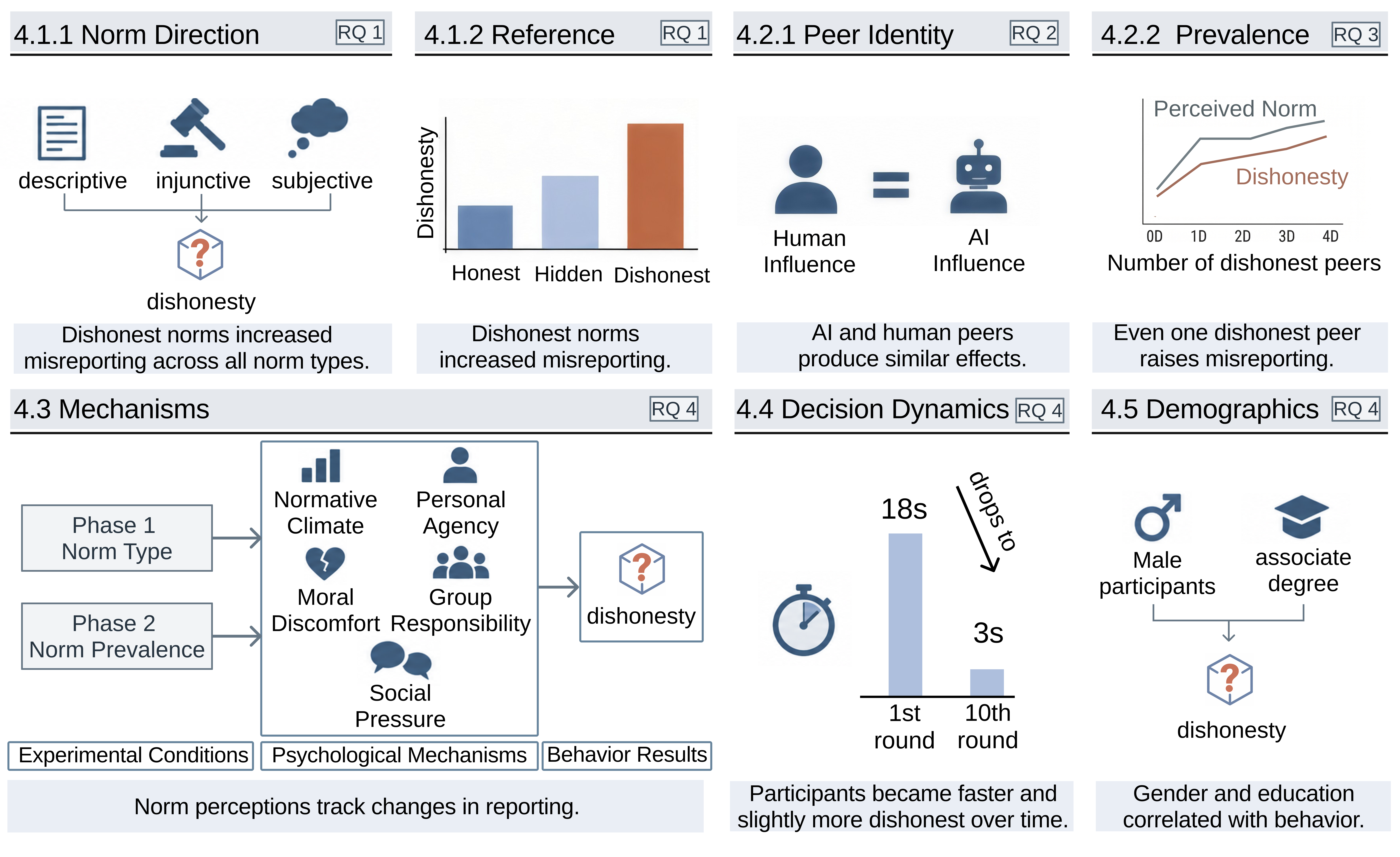}
    \caption{Illustration of this paper's main findings, organized by RQs and subsections.}
    \Description{A visual overview of the paper's main findings. The figure summarizes the effects of social norms, the comparison between human and AI peers, changes associated with dishonest-peer prevalence, candidate psychological mechanisms, repeated-round behavior, and demographic analyses. The central behavioral pattern is that dishonest normative information increases misreporting, while human and AI peer conditions do not show a consistent separation.}
    \label{fig:findings}
\end{figure}

\textbf{The experimental manipulations shifted participants' perceptions of group norms in the intended direction.}
In Phase 1, the \textit{Dishonest} conditions shifted all three direction-sensitive norm measures toward misreporting. The direction effects were significant for descriptive norms ($B=3.15$, 95\% CI [2.76, 3.54], $t(234)=15.92$, $p<.001$, $d=2.06$), injunctive norms ($B=2.63$, 95\% CI [2.27, 2.98], $t(234)=14.39$, $p<.001$, $d=1.85$), and subjective norms ($B=2.80$, 95\% CI [2.45, 3.16], $t(234)=15.59$, $p<.001$, $d=1.97$). In Phase 2, perceived descriptive norms also shifted systematically with dishonest-peer prevalence in the intended direction: each additional dishonest peer reduced the perceived commonness of truthful reporting and increased the perceived commonness of dishonesty (both $p<.001$). The following subsections proceed from the influence of social norms (RQ1), to comparisons between Human and AI peers (RQ2), to the effects of norm prevalence (RQ3), and finally to the psychological mechanisms, repeated-round dynamics, and demographic variation examined in RQ4, as summarized in Figure~\ref{fig:findings}.

\subsection{Influence of Social Norms}

\subsubsection{Norm Direction Across Norm Types}

\begin{figure}[!htbp]
    \includegraphics[width=\textwidth]{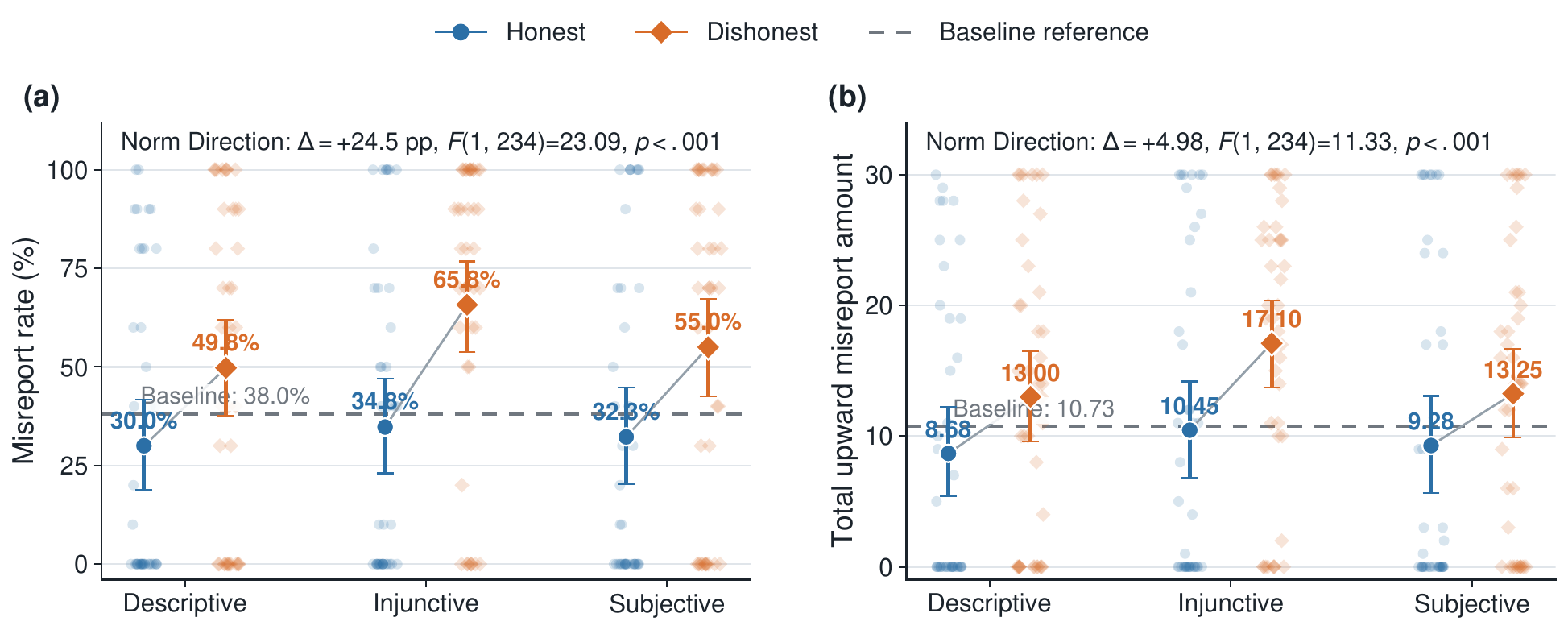}
    \caption{(a) Misreport rate and (b) upward misreport amount across norm types. Error bars indicated 95\% Confidence Interval (CI).}
    \Description{Two participant-level plots compare Honest and Dishonest norm directions across descriptive, injunctive, and subjective norms. In the misreport-rate panel, the Dishonest condition is higher than the Honest condition for all three norm types. Total upward misreport amount shows the same overall directional pattern, although the participant distributions are more variable.}
    \label{fig:social_norm}
\end{figure}

\textbf{\textit{Dishonest} norm conditions significantly increased misreport rates across all three norm types.}
As shown in Figure~\ref{fig:social_norm}, the mean misreport rate increased from 30.0\% to 49.8\% under the descriptive norm condition (Welch $t(77.79)=2.25$, $p=.027$, Cohen's $d=.50$), from 34.8\% to 65.8\% under the injunctive norm condition ($t(77.85)=3.59$, $p<.001$, Cohen's $d=.80$), and from 32.3\% to 55.0\% under the subjective norm condition ($t(77.97)=2.51$, $p=.014$, Cohen's $d=.56$).

\textbf{Across norm types, norm direction significantly affected misreport rate, whereas neither norm type nor its interaction with norm direction was significant.}
The $3\times2$ Type-II ANOVA showed a significant main effect of \textit{Norm Direction} ($F(1,234)=23.09$, $p<.001$, $\eta_p^2=.090$). We did not detect a significant main effect of \textit{Norm Type} ($F(2,234)=1.42$, $p=.245$, $\eta_p^2=.012$), or a significant interaction between \textit{Norm Type} and \textit{Norm Direction} ($F(2,234)=0.44$, $p=.648$, $\eta_p^2=.004$).

\textbf{Upward misreport amount showed the same overall directional tendency, although the within-type contrasts were less consistent.}
Under descriptive norms, the mean upward amount increased from 8.68 to 13.00 (Welch $t(77.99)=1.73$, $p=.089$, Cohen's $d=.39$). Under injunctive norms, it increased from 10.45 to 17.10 ($t(77.11)=2.59$, $p=.012$, Cohen's $d=.58$). Under subjective norms, it increased from 9.28 to 13.25 ($t(77.37)=1.52$, $p=.132$, Cohen's $d=.34$).

\textbf{At the factorial level, norm direction also significantly affected upward misreport amount, whereas neither norm type nor its interaction with norm direction was significant.}
The Type-II ANOVA showed a significant main effect of \textit{Norm Direction} ($F(1,234)=11.33$, $p<.001$, $\eta_p^2=.046$). Neither the main effect of \textit{Norm Type} ($F(2,234)=1.53$, $p=.218$, $\eta_p^2=.013$), nor the interaction between \textit{Norm Type} and \textit{Norm Direction} ($F(2,234)=0.32$, $p=.725$, $\eta_p^2=.003$), was significant. Thus, norm direction significantly affected upward misreport amount overall, although the within-type contrasts were less consistent; we found no significant interaction with norm type.

\subsubsection{Comparison with the Baseline Condition}

\begin{figure}[!htbp]
    \centering
    \includegraphics[width=\linewidth]{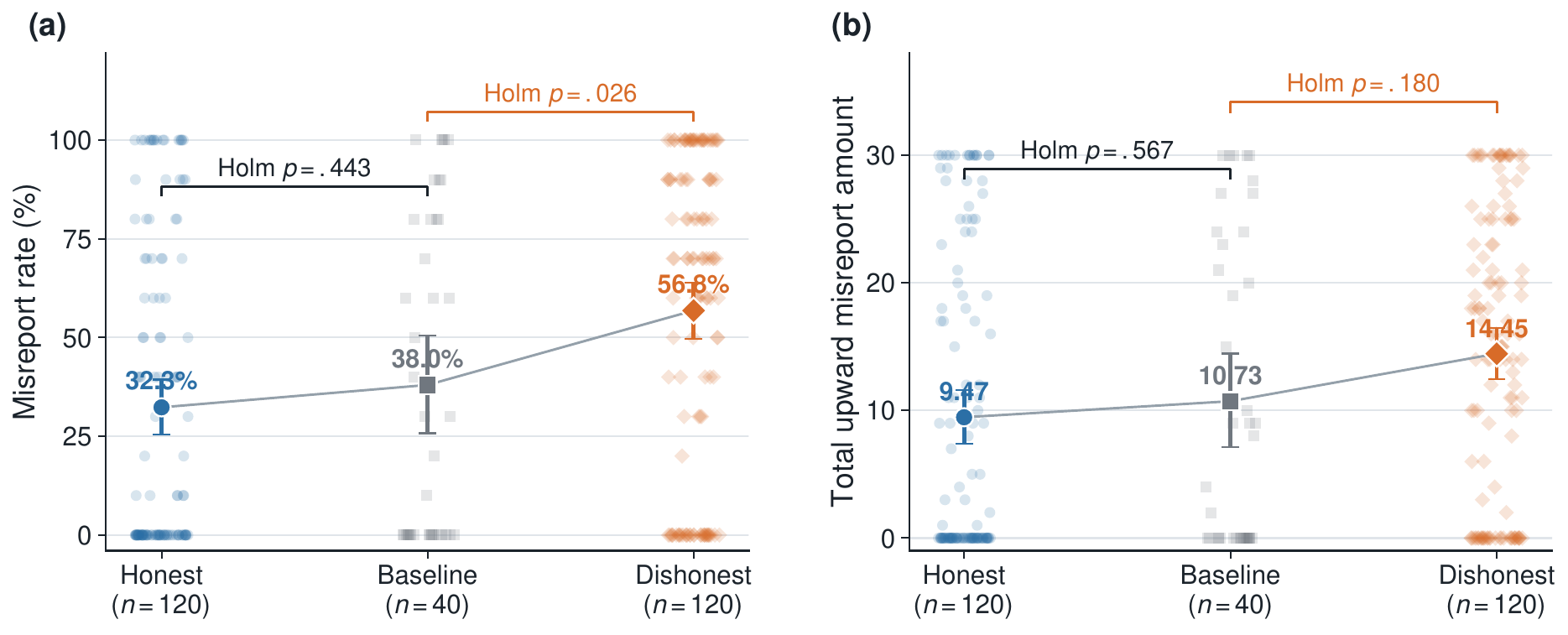}
    \caption{Behavioral outcomes relative to the \textit{Baseline} condition in Phase 1. (a) Misreport rate and (b) total upward misreport amount are shown for the pooled \textit{Honest} conditions, the \textit{Baseline} condition, and the pooled \textit{Dishonest} conditions. Small points represent individual participants, large markers denote condition means, and error bars indicate 20{,}000-resample participant-level bootstrap 95\% confidence intervals.}
    \Description{Two participant-level plots compare pooled Honest conditions, the Baseline condition, and pooled Dishonest conditions. In the misreport-rate panel, the Baseline mean lies between the Honest and Dishonest means, with Dishonest reporting higher than Baseline and Honest reporting slightly lower than Baseline. The upward-misreport-amount panel shows the same ordering but with substantial overlap among participant distributions.}
    \label{fig:invisible_reference}
\end{figure}

\textbf{\textit{Dishonest} norm conditions significantly increased misreport rate relative to \textit{Baseline}, whereas \textit{Honest} norm conditions did not significantly reduce it below \textit{Baseline}.}
As shown in Figure~\ref{fig:invisible_reference}, mean misreport rate was 32.3\% across the \textit{Honest} conditions, 38.0\% in the \textit{Baseline} condition, and 56.8\% across the \textit{Dishonest} conditions. The \textit{Dishonest} conditions showed a significantly higher misreport rate than \textit{Baseline} ($\Delta=18.8$ percentage points, $p_{\mathrm{Holm}}=.026$, Cohen's $d=.47$), whereas the difference between the \textit{Honest} conditions and \textit{Baseline} was not significant ($\Delta=-5.7$ percentage points, $p_{\mathrm{Holm}}=.443$, $d=-.14$).

\textbf{Neither \textit{Honest} nor \textit{Dishonest} norm conditions differed significantly from \textit{Baseline} in total upward misreport amount.}
Neither the \textit{Honest} versus \textit{Baseline} contrast ($p_{\mathrm{Holm}}=.567$) nor the \textit{Dishonest} versus \textit{Baseline} contrast ($p_{\mathrm{Holm}}=.180$) was significant.

\subsection{Peer Identity and Norm Prevalence}

\subsubsection{Effects of Peer Identity}
\textbf{\textit{Norm Prevalence} significantly affected misreport rate, whereas neither \textit{Peer Identity} nor its interaction with \textit{Norm Prevalence} was significant.}
For misreport rate, the $2\times6$ Type-II ANOVA showed a significant main effect of \textit{Norm Prevalence} ($F(5,348)=5.24$, $p<.001$, $\eta_p^2=.070$). We did not detect a significant main effect of \textit{Peer Identity} ($F(1,348)=2.54$, $p=.112$, $\eta_p^2=.007$), or a significant \textit{Peer Identity}$\times$\textit{Norm Prevalence} interaction ($F(5,348)=0.71$, $p=.614$, $\eta_p^2=.010$).

\textbf{The same factorial pattern held for total upward misreport amount.}
\textit{Norm Prevalence} had a significant main effect ($F(5,348)=3.32$, $p=.006$, $\eta_p^2=.046$), whereas neither \textit{Peer Identity} ($F(1,348)=1.07$, $p=.303$, $\eta_p^2=.003$), nor the \textit{Peer Identity}$\times$\textit{Norm Prevalence} interaction ($F(5,348)=0.51$, $p=.772$, $\eta_p^2=.007$) was significant.

\textbf{Pooling across \textit{Norm Prevalence} conditions, Human and AI peers did not differ significantly on either behavioral outcome.}
Across all six \textit{Norm Prevalence} conditions, mean misreport rate was 47.0\% in the Human condition and 40.1\% in the AI condition. The AI--Human difference was $-6.89$ percentage points (95\% CI [-15.61, 1.83], $d=-.16$, $p=.121$). Mean total upward misreport amount was 12.47 in the Human condition and 11.18 in the AI condition ($\Delta=-1.29$, 95\% CI [-3.73, 1.18], $d=-.11$, $p=.309$).

\begin{figure*}[!htbp]
    \centering
    \includegraphics[width=\textwidth]{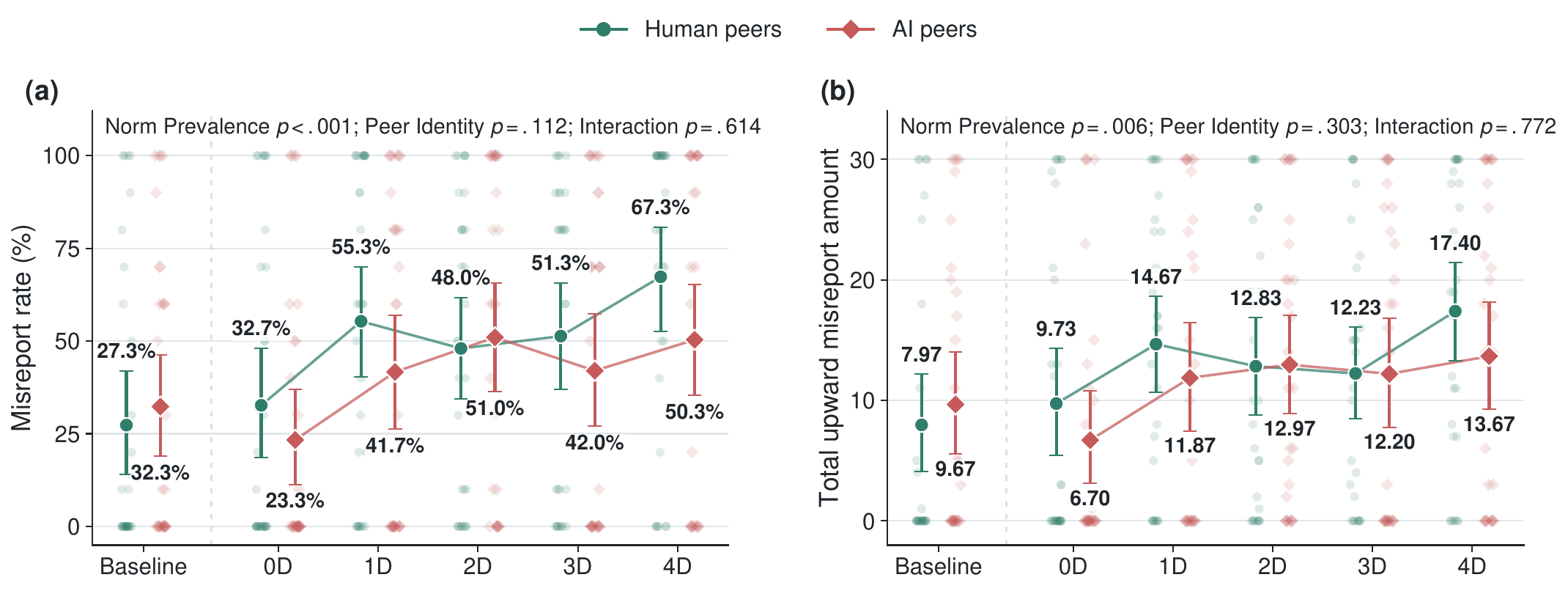}
    \caption{
    Behavioral outcomes across \textit{Peer Identity} and \textit{Norm Prevalence}.
    \textbf{(a)} Participant-level misreport rate and
    \textbf{(b)} total upward misreport amount across the \textit{Baseline} and 0D--4D conditions.
    Circles denote Human peers and diamonds denote AI peers.
    Small points represent individual participants, large markers denote condition means, and error bars indicate 20{,}000-resample participant-level bootstrap 95\% confidence intervals around the means.
    Lines connect the visible 0D--4D conditions. 
    }
    \Description{Two participant-level plots compare Human and AI peers across the Baseline and 0D--4D conditions. Circles represent Human peers and diamonds represent AI peers. The Human and AI patterns overlap substantially across conditions without showing a consistent separation. Dishonest reporting generally increases when dishonest peers are present, but changes from one to four dishonest peers are uneven rather than steadily increasing.}
    \label{fig:human_ai}
\end{figure*}

\subsubsection{Behavioral and Perceived Norm Effects of Norm Prevalence}

\begin{figure*}[!htbp]
    \centering
    \includegraphics[width=\textwidth]{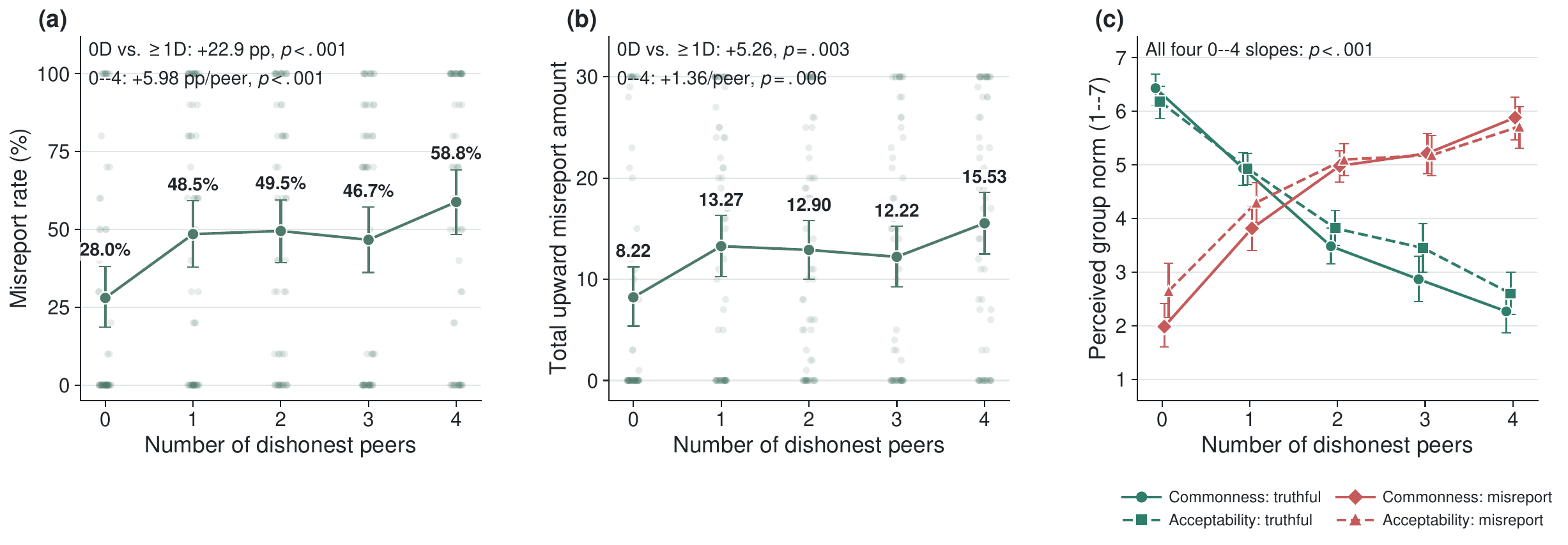}
    \caption{
    Behavioral and perceived norm responses across \textit{Norm Prevalence}, pooled across \textit{Peer Identity}.
    \textbf{(a)} Participant-level misreport rate,
    \textbf{(b)} total upward misreport amount, and
    \textbf{(c)} perceived descriptive norms (commonness) and perceived injunctive norms (acceptability) for truthful reporting and misreporting across the 0D--4D conditions.
    In panels (a)--(b), small points represent individual participants and large markers denote condition means. Panel (c) displays condition means.
    Error bars indicate 20{,}000-resample participant-level bootstrap 95\% confidence intervals around the means.
    In panel (c), color denotes reporting direction and line style denotes perceived descriptive norms (solid) versus perceived injunctive norms (dashed).
    }
    \Description{Three panels show behavioral and perceived norm responses as the number of dishonest peers increases from zero to four. Misreport rate and total upward misreport amount rise sharply when moving from zero to one dishonest peer, then vary non-monotonically across higher prevalence levels. In contrast, perceived descriptive and injunctive norms change progressively with prevalence: truthful reporting is perceived as less common and less acceptable, while misreporting is perceived as more common and more acceptable.}
    \label{fig:prevalence_perception}
\end{figure*}

\textbf{Visible unanimous honesty did not significantly reduce dishonest reporting relative to \textit{Baseline}, whereas exposure to at least one dishonest peer significantly increased both behavioral outcomes relative to 0D.}
We decomposed the \textit{Norm Prevalence} effect to distinguish visible unanimous honesty from the presence of dishonest peer behavior (Fig.~\ref{fig:prevalence_perception}a--b). Mean misreport rates were 29.8\% in \textit{Baseline} and 28.0\% in 0D ($\Delta=-1.83$ percentage points, 95\% CI [$-15.67$, 12.00], $p=.797$). Total upward misreport amounts were 8.82 and 8.22, respectively ($\Delta=-0.60$, 95\% CI [$-4.78$, 3.57], $p=.781$). By contrast, mean misreport rate increased from 28.0\% in 0D to 50.9\% across conditions containing at least one dishonest peer ($\Delta=22.88$ percentage points, 95\% CI [11.54, 33.79], $d=.55$, $p<.001$). Total upward misreport amount similarly increased from 8.22 to 13.48 ($\Delta=5.26$, 95\% CI [1.93, 8.52], $d=.44$, $p=.003$). The corresponding increases in misreport rate were descriptively nearly identical in the Human (+22.83 percentage points) and AI (+22.92 percentage points) conditions; the omnibus \textit{Peer Identity}$\times$\textit{Norm Prevalence} interaction was non-significant, as reported above.

\textbf{Across 0D--4D, greater dishonest-peer prevalence was positively associated with both behavioral outcomes, but no significant linear trend remained once dishonest peer behavior was already present (1D--4D).}
Across the full 0D--4D range, each additional dishonest peer was associated with a 5.98-percentage-point increase in misreport rate ($p<.001$) and a 1.36-point increase in total upward misreport amount ($p=.006$). However, within the restricted 1D--4D range, the linear trends were not significant for either misreport rate ($p=.246$) or total upward misreport amount ($p=.374$). Among the four adjacent comparisons in misreport rate, only the 0D$\rightarrow$1D increase remained significant after Holm correction ($\Delta=20.5$ percentage points, $p_{\mathrm{Holm}}=.041$).

\textbf{Perceived descriptive and injunctive norms shifted progressively with dishonest-peer prevalence, whereas behavioral trends within 1D--4D were non-significant.}
Across the 0D--4D conditions, each additional dishonest peer reduced the perceived \textit{commonness} of truthful reporting by 1.04 points (95\% CI [$-1.16$, $-0.92$], $p<.001$) and increased the perceived commonness of misreporting by 0.92 points (95\% CI [0.79, 1.05], $p<.001$). Perceived \textit{acceptability} showed the same directional pattern: the acceptability of truthful reporting decreased by 0.87 points per additional dishonest peer (95\% CI [$-0.98$, $-0.75$], $p<.001$), whereas the acceptability of misreporting increased by 0.70 points (95\% CI [0.56, 0.84], $p<.001$).

\subsection{Underlying Mechanisms}

We further explored through what psychological mechanisms participants were influenced by examining possible modulating mechanisms. Using the exit questionnaire results, Structural Equation Models (SEM) were estimated with maximum-likelihood Wishart estimation (MLW). We report model chi-square, CFI, TLI, RMSEA (with 90\% confidence intervals), SRMR, AIC, BIC, standardized path coefficients, scale reliability and candidate indirect effects. 

For the Phase 1 SEM, descriptive, injunctive, and subjective norm perceptions were modeled as related factors of a normative perception construct. \textit{Perceived social pressure}, \textit{moral discomfort}, \textit{personal agency}, and \textit{shared group responsibility} were specified as post-task mechanisms, while the seven experimental conditions were entered as predictors. The model showed good fit, $\chi^2(43)=57.83$, $p=.065$, $\chi^2/df=1.34$, CFI=.989, TLI=.976, RMSEA=.035, SRMR=.018, AIC=123.59, and BIC=348.94. 

\textbf{Dishonest normative cues and the absence of visible peer reports increased perceived dishonest norms, while exerting negative effects on personal agency and social pressure.} The three normative indicators loaded strongly onto the norm construct: descriptive ($\beta=.920$), injunctive ($\beta=.913$), and subjective norms ($\beta=.874$, all $p<.001$).
Relative to the honest condition with injunctive norm, perceived normative support for dishonesty was heightened across all dishonest conditions (injunctive: $\beta=.640$, $p<.001$; subjective: $\beta=.598$, $p<.001$; descriptive: $\beta=.544$, $p<.001$) as well as in the baseline condition ($\beta=.246$, $p<.001$). Conversely, the honest condition with subjective norm ($\beta=.074$, $p=.185$) and honest condition with descriptive norm ($\beta=.032$, $p=.571$) did not differ significantly from the baseline condition. Furthermore, exposure to dishonest injunctive cues significantly diminished personal agency ($\beta=-.251$, $p<.001$), while the honest condition with descriptive norm ($\beta=-.201$, $p=.004$) and the baseline condition ($\beta=-.353$, $p<.001$) were associated with lower perceived social pressure. Experimental conditions had no significant effects on moral discomfort or shared group responsibility.

\textbf{Dishonest behavior is affected by heightened normative support, elevated moral discomfort, and attenuated shared responsibility.} Normative perception positively predicted misreporting behavior ($\beta=.237$, $p=.009$). Moral discomfort was also positively associated with misreporting ($\beta=.273$, $p<.001$), while shared group responsibility was negatively associated with misreporting ($\beta=-.171$, $p=.003$). Perceived social pressure ($\beta=-.103$, $p=.148$) and personal agency ($\beta=-.035$, $p=.580$) were not significant predictors. None of the conditions retained a significant direct association with misreporting behavior after accounting for psychological mechanisms.

\textbf{Mediation analyses found that perceived normative support is the psychological mechanism transmitting the effects of dishonest peer cues onto misreporting behavior.} Bootstrap analyses indicated significant indirect effects through normative perception for the dishonest condition with injunctive norms ($b=.156$, 95\% CI [.028, .315], $p=.016$), the dishonest condition with subjective norms ($b=.145$, 95\% CI [.027, .271], $p=.016$), the dishonest condition with descriptive norms ($b=.132$, 95\% CI [.025, .241], $p=.016$), and the baseline condition ($b=.060$, 95\% CI [.011, .118], $p=.016$). The indirect effect for the honest condition with subjective norms was not significant ($b=.019$, [-.010, .060], $p= .204$), and neither was the indirect effect for the honest condition with descriptive norms ($b=.007$, [-.021, .045], $p=.548$). None of the indirect effects through perceived social pressure, moral discomfort, personal agency, or shared group responsibility was significant (see Appendix~\ref{sec:indirect_nonsignificant} for reports).

\begin{figure}[!htbp]
    \centering
    \includegraphics[width=\textwidth]{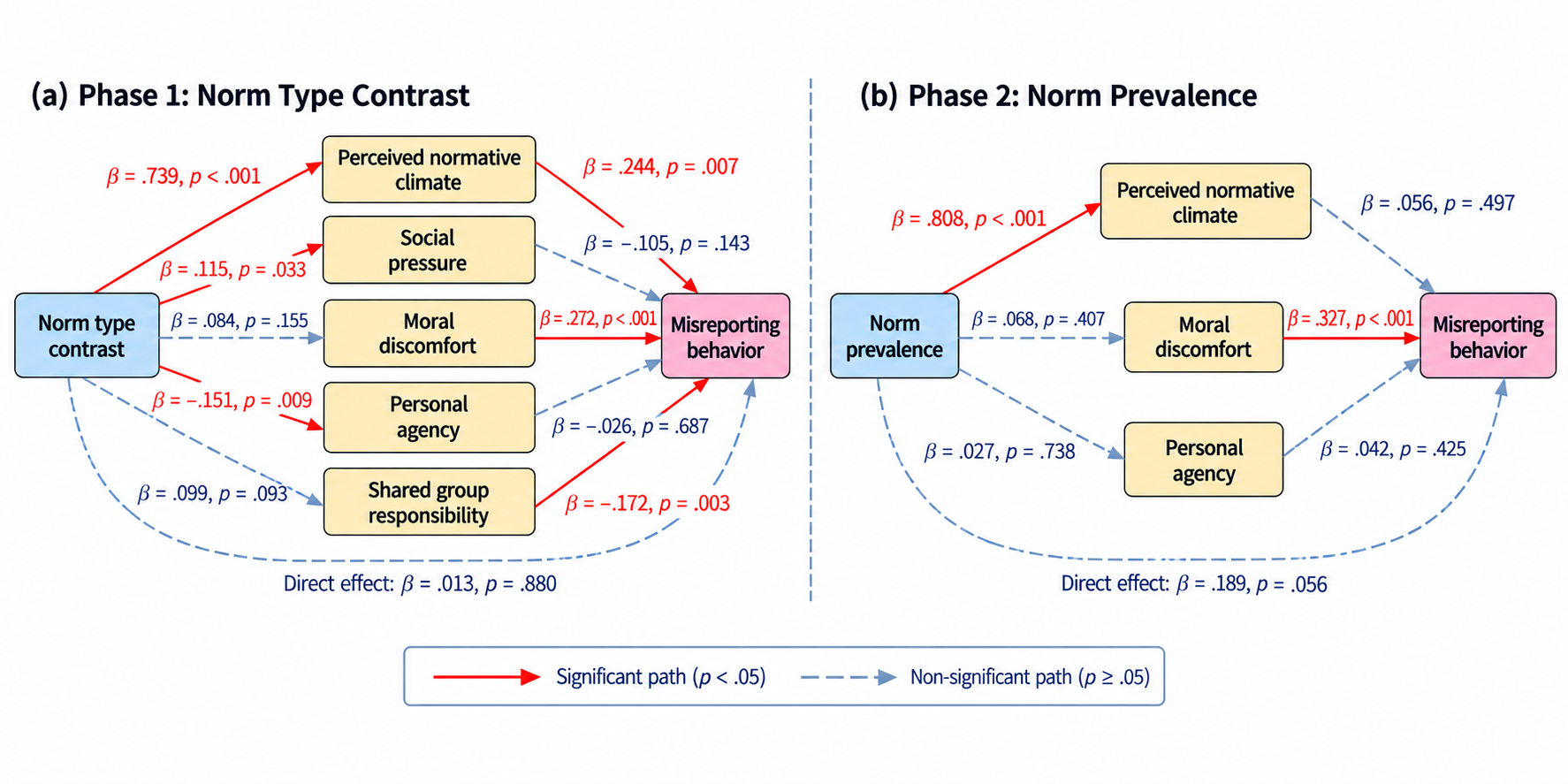}
    \Description{Two side-by-side path diagrams illustrate the proposed psychological pathways in Phase 1 and Phase 2. The left panel shows a norm contrast connected to five candidate mechanisms—perceived normative climate, social pressure, moral discomfort, personal agency, and shared group responsibility—before reaching misreporting behavior. The right panel shows norm prevalence connected to perceived normative climate, moral discomfort, and personal agency before reaching misreporting behavior. Significant paths are highlighted with solid red arrows, non-significant paths with gray dashed arrows, and each arrow is annotated with its standardized coefficient and p value.}
    \caption{The SEM for (a) Phase 1 and (b) Phase 2.}
\end{figure}

In Phase 2, \textit{Norm Prevalence} was modeled as a predictor of perceived misreport norms. Moral discomfort and personal agency were included as parallel candidate mechanisms to account for post-decision moral evaluation and altered ownership, assuming no directional paths among these mechanisms. The model showed acceptable approximate fit: $\chi^2(52)=126.51$, $p<.001$, $\chi^2/df=2.43$, CFI=.970, TLI=.952, RMSEA=.063 (90\% CI [.049, .077]), SRMR=.037, AIC=77.30, and BIC=228.86.

\textbf{Perceptions of a permissive normative climate were driven by dishonesty prevalence and the absence of peer reports.} Peer dishonesty prevalence is correlated with a dishonest normative climate ($\beta=.808$, $z=13.00$, $p<.001$), confirming that observing widespread dishonesty legitimized misreporting. Similarly, relative to the visible honest condition, assignment to the baseline condition strengthened perceived dishonest norms ($\beta=.347$, $z=7.58$, $p<.001$), implying that the absence of observable peer information is construed as more permissive of dishonesty than explicit honesty. Normative climate was unaffected by peer identity ($\beta=-.078$, $z=-1.33$, $p=.184$), or the prevalence-by-AI identity interaction ($\beta=.084$, $z=1.20$, $p=.232$), suggesting that human and AI peers influenced normative perceptions similarly. Furthermore, observing progressively more dishonest behavior did not significantly alter participants' subsequent moral discomfort ($\beta=.068$, $z=0.83$, $p=.407$) or personal agency ($\beta=.027$, $z=0.33$, $p=.738$). 

\textbf{Misreporting in Phase 2 was primarily linked to moral discomfort and direct prevalence exposure, with no evidence of behavioral differentiation between human and AI peers.} Controlling for candidate mechanisms and experimental predictors, the conditional direct path from norm prevalence to misreport behavior was marginally significant ($\beta=.189$, $z=1.91$, $p=.056$). Consistent with Phase 1, moral discomfort is independently associated with more frequent misreporting ($\beta=.327$, $z=6.02$, $p<.001$), reinforcing the interpretation that moral discomfort acts as a post-task consequence of dishonest behavior. By contrast, perceived normative climate ($\beta=.056$, $z=0.68$, $p=.497$) and personal agency ($\beta=.042$, $z=0.80$, $p=.425$) did not independently explain additional behavioral variation. AI peer identity ($\beta=-.001$, $z=0.02$, $p=.987$) and its interaction with prevalence ($\beta=-.085$, $z=-0.97$, $p=.333$) showed no association with behavior. Collectively, these results provide no evidence that behavioral responses differed by \textit{Peer Identity}.

\textbf{Candidate indirect pathways failed to account for the link between peer dishonesty and participant misreporting.} Indirect effect analysis identified no significant indirect effects linking peer dishonesty to misreporting through perceived normative climate ($b=.014$, 95\% CI [-.026, .054]), moral discomfort ($b=.006$, 95\% CI [-.008, .021]), or personal agency ($b<.001$, 95\% CI [-.002, .004]).

\subsection{Learning Effects}\label{sec:learning_effects}

\textbf{Decision times decreased sharply across repeated reporting rounds, indicating rapid task familiarization.}
Median decision time decreased from 18.13 seconds in Round~1 to 6.47 seconds in Round~2 and 2.93 seconds in Round~10. In the piecewise GEE analysis, decision time decreased by approximately 57.2\% from Round~1 to Round~2 (95\% CI [52.9\%, 61.2\%], $p<.001$) and continued to decrease by approximately 8.2\% per round across Rounds~2--10 (95\% CI [7.2\%, 9.1\%], $p<.001$).

\textbf{From Round~1 to Round~2, misreport odds increased significantly within the \textit{Dishonest} conditions, but the early trajectories did not differ significantly by norm direction.}
Misreport odds increased in the \textit{Dishonest} conditions ($\mathrm{OR}=1.39$, 95\% CI [1.01, 1.92], $p=.044$), whereas no corresponding change was detected in the \textit{Honest} conditions ($\mathrm{OR}=1.01$, 95\% CI [0.73, 1.40], $p=.942$). However, the \textit{Norm Direction}$\times$\textit{Round} interaction was not significant ($p=.139$), providing insufficient evidence that the early change differed between the \textit{Honest} and \textit{Dishonest} conditions. For total upward misreport amount, the increase within the \textit{Dishonest} conditions was not significant ($B=.21$, 95\% CI [-.02, .44], $p=.078$), and the corresponding \textit{Norm Direction}$\times$\textit{Round} interaction was also non-significant ($p=.071$).

\textbf{From Rounds~2--10, both behavioral outcomes increased modestly over time, with no significant moderation by norm direction.}
The odds of misreporting increased by approximately 3.1\% per round ($\mathrm{OR}=1.031$, 95\% CI [1.006, 1.056], $p=.014$), while total upward misreport amount increased by $B=.028$ points per round (95\% CI [.010, .045], $p=.002$). Neither later trajectory differed significantly by \textit{Norm Direction} (misreport incidence: $p=.978$; total upward misreport amount: $p=.068$).

Overall, participants became faster with repeated task exposure and showed modest increases in dishonest reporting over later rounds. However, these temporal changes did not significantly differ between the \textit{Honest} and \textit{Dishonest} conditions.


\subsection{Effects of Demographics}

\begin{figure}[!htbp]
    \centering 
    \subfloat[Regression analysis for Phase 1.]{
        \includegraphics[width=0.5\textwidth]{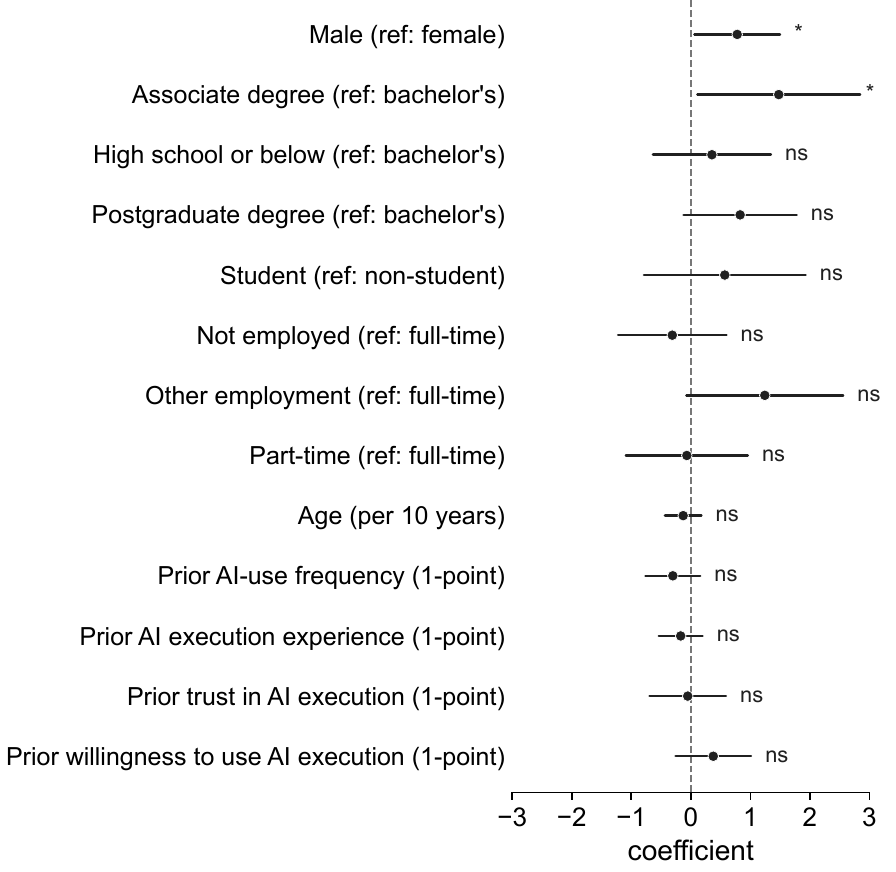}
    }
    \subfloat[Regression analysis for Phase 2.]{
        \includegraphics[width=0.5\textwidth]{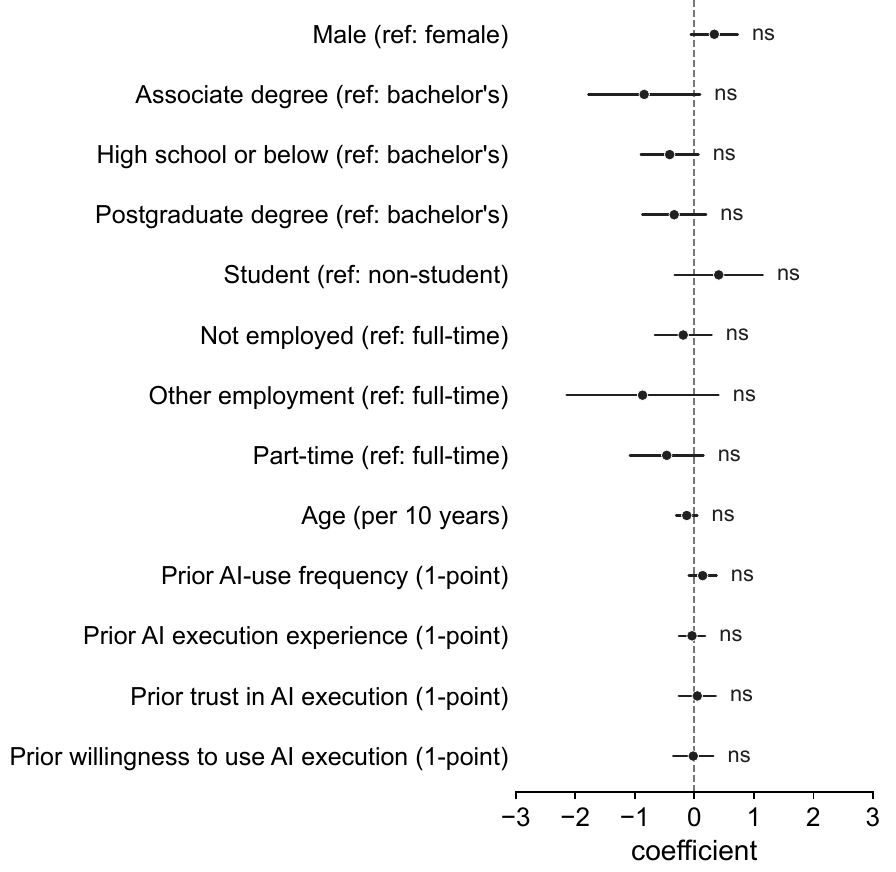}
    }
    \caption{The regression analyses for (a) Phase 1, and (b) Phase 2. Error bar indicated 95\% CI for the coefficient estimate. ns, * denotes $p \ge .05$, $p < .05$ respectively.}
    \Description{Two forest plots show regression estimates for demographic characteristics and prior AI-related dispositions in Phase~1 and Phase~2. Points indicate coefficient estimates and horizontal intervals indicate 95\% confidence intervals, with a vertical reference line at zero representing no association. Most estimates are close to zero or have confidence intervals crossing zero, while a small number of demographic contrasts show statistically significant associations with reporting behavior.}
    \label{fig:demographics}
\end{figure}

\textbf{As shown in Figure~\ref{fig:demographics}, in Phase 1, individual behavior variance was associated with gender and specific educational attainment, while other demographic backgrounds and AI-related dispositions show no effects.} Specifically, male participants exhibited significantly higher behavioral scores than female participants ($\beta=0.779$, 95\% CI [0.060, 1.498], $p = .034$). Regarding education, participants holding an associate degree reported higher scores than those with a bachelor's degree ($\beta=1.477$, 95\% CI [0.111, 2.844], $p = .034$), although contrasts for other educational levels remained non-significant (high school or below: $p = .483$; post-graduate: $p = .087$). In contrast, participant behavior was not significantly associated with student status ($p = .410$), employment categories ($p \ge .064$) or age ($p = .412$). Dispositions toward AI, including prior use frequency, task execution experience, trust and delegation willingness, were also not associated with participants' reporting behavior ($p \ge .198$, see Appendix~\ref{app:demographic_models}) for complete parameter estimates.

\textbf{In Phase 2, behavioral variation was robust across socioeconomic status or prior AI attitudes, which suggest that the social influence is prevalent.} Neither gender ($p = .095$), educational status ($p = .277$), nor age ($p = .163$) reached significance. Replicating Phase 1, dispositions towards AI all show no significant associations with reporting behavior ($p \ge .238$, detailed in Appendix~\ref{app:demographic_models}).

\section{Discussions}

\subsection{AI Agents and Their Social Influence}

Our findings advance HCI community's understanding of unethical conduct in mixed human-AI teams. We contextualize our contributions from AI's social influence to its interpretation, and further to the characteristics of such influence. 

\textbf{First, we advance the conceptualization of AI in ethical decision making by showing its efficacy as an active social influencer.} Prior frameworks, such as Kobis et al.~\cite{kobis2021badmachines,kobis2025delegation}, primarily examined AI as an enabler by showing how humans delegate dishonest tasks to automated systems. Our studies address the role of AI as an influencer and an advisor. We show that AI agents acting as group peers directly shape human choices through social norms. Across descriptive, injunctive, and subjective norm conditions, AI agents communicate standards of behavior that participants actively follow. In descriptive settings, as an influencer, AI peers show specific reporting actions that humans reproduce. In injunctive and subjective settings, as an advisor, AI peers state what is acceptable or expected, creating social expectations that reduce personal responsibility, which reasonates with recent findings that LLMs can elicit private disclosures~\cite{zhang2025power}. While prior research tested the effect of AI advice in two-party interactions~\cite{leib2024corrupted,kim2024robotadvisers}, we extend the dyad interaction scope by showing that AI agents operating within groups directly alter human reporting behavior. 

\textbf{Second, our empirical results resolve conflicting theoretical expectations regarding the comparisons of AI influence and human social influence.} One perspective, based on Panoptic Surveillance theory~\cite{foucault2012discipline,weng2026ai} and social control mechanisms~\cite{kroher2015socialcontrol}, suggests that people view AI as a monitoring tool, which would automatically decrease dishonest behavior. Another perspective, derived from Automation Bias~\cite{mosier1998automation} and Algorithm Appreciation~\cite{logg2019algorithm}, suggests that people follow AI decisions much more closely than human decisions. Our data do not support either interpretation. AI peers did not reduce dishonesty across all conditions, nor did they exert a stronger influence than human peers. Instead, misreport rates in the AI condition were statistically indistinguishable from those in the human condition. These results extend Social Influence theory~\cite{turner1991social} and CASA paradigm~\cite{nass1994computers}. While prior CASA studies showed that people apply social rules to computers in non-incentivized tasks~\cite{song2025multi,feng2025multi}, our findings show that this social influence persists in incentivized situations where following dishonest AI behavior produces direct financial gains. 

\textbf{Third, we contribute empirical insights on minority influence, showing how group composition of AIs affect ethical conducts.} Standard conformity models~\cite{asch1956studies,latane1981psychology} assume that conformity increases as more group members adopt a behavior. In our study, conformity did not increase linearly with the number of dishonest peers. Instead, shifting from zero dishonest peers to one dishonest peer caused the primary increase, while adding subsequent dishonest peers produced minimal further change. This pattern aligns with the minority influence research~\cite{moscovici1969influence}. In financial reporting tasks, individuals balance monetary incentives against the psychological cost of lying~\cite{abeler2019preferences,mazar2008dishonesty}. Observing a single dishonest peer signals that rule violations occur in the group, which gives individuals an external justification to misreport for profit~\cite{shalvi2015self}. Because the initial dishonest peer establishes that misreporting is evident, additional dishonest peers do not substantially lower the threshold for dishonesty.

\textbf{Furthermore, we identify a divergence between participants' continuous normative perceptions and their non-linear behavioral responses.} While perceived commonness and acceptability shifted progressively with each additional dishonest peer, behavioral responses across 1D--4D were substantially less monotonic. This divergence suggests that people may track changes in a group's normative composition more precisely than they translate those changes into actions. Perceiving a group as increasingly permissive of dishonesty does not guarantee an equally incremental escalation in misconduct, probably because translating perception into action requires overcoming internal moral friction. This disconnect between acknowledged normative shifts and actual conduct mirrors the ``privacy paradox''~\cite{kokolakis2017privacy}, extending the concept of attitude-behavior misalignment into ethical conducts.

\textbf{Fourth, we contribute interpretations on why AI misconduct effectively erodes human moral accountability through Mind Perception theory~\cite{gray2007mind}.} When interacting with other humans, prior work suggested that individuals consider shared relational accountability and mutual moral standards~\cite{weisel2015collaborative}. In contrast, individuals attribute agency to AI agents but do not attribute subjective moral feelings, such as guilt, to them~\cite{gray2007mind,shank2025moral}. Our studies confirmed such interpretations in unethical conduct settings. Our participants observed an AI peer misreport without facing human moral judgment or group social disapproval. As reflected in participants' subjective rating, they treated the AI's misreport as an objective indicator that misreporting was possible within the task rules. This reduced the individual's perceived accountability, which corresponds with our SEM showing that exposure to dishonest AI signals significantly decreased feelings of personal agency~\cite{zhang2025exploring}. 

\textbf{Finally, our decision time data reveal exploratory insights on how repeated exposure may alter decision structures of human ethical choices.} Interpreted through the lens of Dual-Process theory, people rely on deliberate cognitive reasoning (System 2) or fast heuristic processing (System 1)~\cite{kahneman2011thinking,shalvi2012honesty}. In the first round, participants took an average of 18.13 seconds to submit their report, indicating deliberate moral reasoning about whether to report truthfully. In later rounds, median decision time dropped below 3 seconds. This reduction in time suggests that participants may have stopped deliberating about each individual choice. Repeated exposure to the group environment may convert the reporting task into an automatic process, which confirmed prior empirical findings on how individuals habituate to recurring ethical transgressions over time~\cite{gino2009misconduct,welsh2015slippery}.

\subsection{Demographic Heterogeneity and Socio-Cultural Implications}

While the primary findings underscore the social influence of AI agents on human dishonesty, our demographics analyses reveal heterogeneities in how different subgroups respond to AI-mediated normative cues. These variations indicate that human-AI moral contagion is situated within individual and socio-cultural contexts. In Phase 1, gender and educational attainment are significant covariates. Male participants exhibited a greater propensity to align with dishonest norms, mirroring behavioral economics literature linking men to higher risk tolerance and susceptibility to competitive moral transgressions~\cite{croson2009gender}. Furthermore, behavioral patterns across educational states suggest that formal education influences individual moral frameworks or the interpretation of experimental incentives. Although we did not identify other correlating demographic factors in both Phase 1 and Phase 2, ethical choices may still be subject to cultural or region-specific norms. For example, institutional trust, cultural tightness, or localized exposure to algorithmic management may modulate the perceived legitimacy of AI peers~\cite{gelfand2011differences,kellogg2020algorithms}. Similarly, perceived surveillance, accountability, and institutional trust may influence users' behavior in groups, especially when faced with ethical decisions~\cite{brayne2017big,burrell2016machine}. Therefore, whether the results could be transferred to other regions and contexts, such as the Eastern context, requires further empirical validation.

\subsection{Implications}

We structure our recommendations from both AI-side and human-side:

\textbf{[AI] Setting verified behavioral baselines and isolating non-compliant agents.} Our findings demonstrate that visible unanimous honesty did not significantly reduce dishonest reporting relative to the \textit{Baseline} condition, whereas exposure to dishonest peer behavior was associated with substantially greater misreporting. Crucially, the introduction of a single non-compliant AI agent is sufficient to trigger a large escalation in human dishonesty, matching the influence of human peers  with diminishing marginal effects from additional dishonest agents. Consequently, systems should actively display verified, truthful AI baselines rather than relying on informational opacity. Furthermore, multi-agent architectures should implement real-time anomaly detection to isolate deviant agents immediately, as even isolated instances of non-compliance can compromise collective integrity.

\textbf{[AI] Establishing multi-dimensional normative alignment and equal compliance standards.} Participants exhibited comparable behavioral shifts whether exposed to descriptive norms, injunctive norms, or subjective norms. This indicates that social influence operates through multiple communication channels rather than observable behavior alone. Because individuals respond to artificial peers with the same behavioral susceptibility as human peers in financially incentivized settings, artificial agents cannot be treated as passive computational tools. Regulatory frameworks and organizational compliance standards must hold AI agents to the same ethical oversight, communicative boundaries, and auditing protocols as human personnel, explicitly constraining communicative outputs to prevent the transmission of dishonest injunctive and subjective cues.

\textbf{[Human] Disrupting automated conformity through dynamic cognitive friction.} Decision latency decreased sharply from 18.13 seconds in early trials to under 3 seconds in later trials, while dishonest reporting increased across repeated rounds. This pattern reveals that sustained exposure to routine workflows transitions ethical decision-making from deliberate cognitive processing to automated heuristics. To prevent this habituation, interfaces should introduce targeted cognitive friction—such as mandatory verification pauses or active justification prompts—specifically when users execute rapid, repetitive actions in environments susceptible to misreporting.

\textbf{[Human] Enforcing personal agency with demographic-sensitive scaffolding.} Exposure to peer non-compliance significantly reduced participants' perceived decision ownership, indicating a diffusion of responsibility in collaborative groups. To re-engage individual accountability, interfaces should incorporate explicit authorship mechanisms, such as mandatory confirmation steps that assign singular responsibility to the human operator prior to submission. Furthermore, because susceptibility to normative cues varied across socio-demographic contexts (e.g., geographic location), these interface safeguards should be calibrated to specific user cohorts and organizational norms rather than implemented as uniform defaults.

\section{Limitations and Future Work}

We acknowledge several limitations of this paper. First, while our use of established incentivized reporting paradigms ensures high internal validity, real-world workplace settings involve higher financial stakes, long-term relational dynamics, and complex organizational hierarchies. Future work could evaluate whether such social influence effect extends in naturalistic organizational settings.

Second, in our experiment, AI agents were presented via structured interface elements to control for confounding variables such as anthropomorphism and conversational style. Future work could introduce human-like avatars or natural language interaction to assess their moderating effect on social influence and normative effects.  

Third, our repeated measures design captured the early-stage dynamics and convergence of normative decision making. However, future studies should examine whether extended periods would cause user habituation to AI agents, or alters the observed social influence over time. 

Finally, we held variables such as group size constant to prioritize primary hypotheses. However, future studies should expand on this framework by varying team size, structure and composition to map the broader boundary conditions of social influence by AI peers. 

\section{Ethical Considerations}

We acknowledge ethical implications of this paper, and followed the Menlo report~\cite{bailey2012menlo} and Belmont report~\cite{beauchamp2008belmont} in mitigating risks. All the two phases got the approval of XX\footnote{anonymized for submission} university's IRB.

\textbf{Respect for persons.} In our study, participants were informed of their rights prior to participation, including the right to withdraw from the study at any point. We used an incomplete disclosure protocol, where we disclosed the experiment settings before the study, and provided debriefing statement immediately upon study completion, fully disclosing the artificial nature of peer behavior, providing options for them to quit the study and delete the data, explaining the necessity of the wizard-of-oz design, and offering contact information for further inquiries. No one chose to quit the study and delete their data. 

\textbf{Beneficence.} The experimental tasks were designed as low-stakes, stylized economic decision scenarios that posed minimal psychological or behavioral risk. No sensitive personal identifiers and illegal activities were involved. All participant data were anonymized, stored securely, and processed in aggregate to protect individual privacy and prevent potential social or reputational harm. The controlled use of incomplete disclosure was essential to capture natural behaviors without introducing confounding variables or subjecting participants to real-world operational risks.

\textbf{Justice.} Participants were recruited via Prolific following standardized eligibility criteria. All participants received fair monetary compensation according to the Prolific's recommendation. 

\textbf{Respect for law and public interest.} While our empirical results show that AI peer presence effectively suppresses human misconduct, we caution against the covert or deceptive deployment of artificial agents as surveillance mechanisms in real-world organizations. System designers and policymakers should ensure AI peers respect user trust and individual agency.

\section{Conclusion}

In this paper, we show that AI agents serving as collaborative group peers significantly influence human dishonest behavior in incentivized environments. Across descriptive, injunctive, and subjective norms, exposure to dishonest AI peers substantially increases human misreporting, showing an influence comparable in magnitude to that of human peers. Notably, conformity does not scale linearly, where the introduction of a single dishonest peer is sufficient to trigger the primary surge in misconduct, providing empirical support for a minority influence effect. Moreover, observing peer misreporting attenuates perceived decision ownership, while repeated interactions transition ethical choices from deliberation into automated heuristic processing. These findings extend Social Influence Theory and the CASA paradigm to incentivized ethical conduct. We recommend multi-agent systems enforce strict normative alignment, isolate deviant agents, and integrate interface friction to maintain human agency. 




\bibliographystyle{ACM-Reference-Format}
\bibliography{sample-base}

\appendix

\section{Generative AI Usage}

We used ChatGPT and Gemini-3.1-pro for polishing the text of this paper, including checking grammar. We also used ChatGPT for refining the figures. All authors take full responsibility for all this paper's content.

\section{Demographics}
\label{app:demographics}

Tables~\ref{tbl:demographics_core}--\ref{tbl:demographics_status} summarize the demographic and profile characteristics of the final analyzed samples. Age, sex, ethnicity, geographic profile, language, student status, and employment status were obtained from participants' Prolific profile data. Education was collected in the post-study questionnaire.

\begin{table*}[!htbp]
\centering
\caption{Core demographic characteristics of the final analyzed samples.}
\label{tbl:demographics_core}
\scriptsize
\setlength{\tabcolsep}{4pt}
\renewcommand{\arraystretch}{1.15}
\begin{tabularx}{\textwidth}{p{0.17\textwidth} c p{0.17\textwidth} X X}
\toprule
Sample & $N$ & Age, $M$ ($SD$), range & Sex & Education \\
\midrule

Phase 1: Norm Experiment
& 280
& 39.25 (12.25), 18--81
& Female 131 (46.8\%); Male 146 (52.1\%); Unavailable 3 (1.1\%)
& High school or below 49 (17.5\%); College/associate 32 (11.4\%); Bachelor's 119 (42.5\%); Master's 65 (23.2\%); Doctorate 11 (3.9\%); Other 2 (0.7\%); Prefer not to say 2 (0.7\%) \\

Phase 2: Human peers
& 180
& 40.46 (11.24), 21--68
& Female 93 (51.7\%); Male 86 (47.8\%); Unavailable 1 (0.6\%)
& High school or below 45 (25.0\%); College/associate 11 (6.1\%); Bachelor's 82 (45.6\%); Master's 30 (16.7\%); Doctorate 7 (3.9\%); Other 3 (1.7\%); Prefer not to say 2 (1.1\%) \\

Phase 2: AI peers
& 180
& 37.14 (12.00), 18--79
& Female 105 (58.3\%); Male 72 (40.0\%); Prefer not to say 1 (0.6\%); Unavailable 2 (1.1\%)
& High school or below 49 (27.2\%); College/associate 17 (9.4\%); Bachelor's 76 (42.2\%); Master's 29 (16.1\%); Doctorate 5 (2.8\%); Other 3 (1.7\%); Prefer not to say 1 (0.6\%) \\

\bottomrule
\end{tabularx}
\vspace{2pt}
\begin{minipage}{0.98\textwidth}
\footnotesize
\textit{Note.} Age and sex were obtained from Prolific profile data; education was collected in the post-study questionnaire. Age statistics use available values only (Phase~1: $n=277$; Phase~2 Human peers: $n=179$; Phase~2 AI peers: $n=180$). ``Unavailable'' denotes missing Prolific profile information and was not recoded as ``Prefer not to say.'' Percentages use the full analyzed sample size within each row.
\end{minipage}
\end{table*}

\begin{table*}[!htbp]
\centering
\caption{Additional Prolific profile characteristics: ethnicity and geographic background.}
\label{tbl:demographics_profile}
\scriptsize
\setlength{\tabcolsep}{4pt}
\renewcommand{\arraystretch}{1.15}
\begin{tabularx}{\textwidth}{p{0.17\textwidth} X X X X}
\toprule
Sample & Ethnicity & Country of residence & Country of birth & Nationality \\
\midrule

Phase 1: Norm Experiment
& White 167 (59.6\%); Black 48 (17.1\%); Asian 37 (13.2\%); Mixed 14 (5.0\%); Other 7 (2.5\%); Prefer not to say 3 (1.1\%); Unavailable 4 (1.4\%)
& United Kingdom 197 (70.4\%); United States 80 (28.6\%); Unavailable 3 (1.1\%)
& United Kingdom 137 (48.9\%); United States 70 (25.0\%); Other 70 (25.0\%); Unavailable 3 (1.1\%)
& United Kingdom 147 (52.5\%); United States 75 (26.8\%); Other 51 (18.2\%); Unavailable 7 (2.5\%) \\

Phase 2: Human peers
& White 124 (68.9\%); Black 23 (12.8\%); Asian 18 (10.0\%); Mixed 8 (4.4\%); Other 5 (2.8\%); Prefer not to say 1 (0.6\%); Unavailable 1 (0.6\%)
& United Kingdom 128 (71.1\%); United States 51 (28.3\%); Unavailable 1 (0.6\%)
& United Kingdom 98 (54.4\%); United States 39 (21.7\%); Other 42 (23.3\%); Unavailable 1 (0.6\%)
& United Kingdom 106 (58.9\%); United States 45 (25.0\%); Other 26 (14.4\%); Unavailable 3 (1.7\%) \\

Phase 2: AI peers
& White 123 (68.3\%); Black 17 (9.4\%); Asian 17 (9.4\%); Mixed 13 (7.2\%); Other 9 (5.0\%); Unavailable 1 (0.6\%)
& United Kingdom 65 (36.1\%); United States 115 (63.9\%)
& United Kingdom 52 (28.9\%); United States 98 (54.4\%); Other 30 (16.7\%)
& United Kingdom 53 (29.4\%); United States 105 (58.3\%); Other 17 (9.4\%); Unavailable 5 (2.8\%) \\

\bottomrule
\end{tabularx}
\vspace{2pt}
\begin{minipage}{0.98\textwidth}
\footnotesize
\textit{Note.} All variables in this table were obtained from Prolific profile data. For country of birth and nationality, ``Other'' aggregates all categories other than the United Kingdom and the United States. Percentages use the full analyzed sample size within each row.
\end{minipage}
\end{table*}

\begin{table*}[!htbp]
\centering
\caption{Additional Prolific profile characteristics: language and socioeconomic status.}
\label{tbl:demographics_status}
\scriptsize
\setlength{\tabcolsep}{4pt}
\renewcommand{\arraystretch}{1.15}
\begin{tabularx}{\textwidth}{p{0.17\textwidth} X X X}
\toprule
Sample & Language & Student status & Employment status \\
\midrule

Phase 1: Norm Experiment
& English 227 (81.1\%); Other 49 (17.5\%); Unavailable 4 (1.4\%)
& No 211 (75.4\%); Yes 38 (13.6\%); Unavailable 31 (11.1\%)
& Full-time 140 (50.0\%); Part-time 45 (16.1\%); Unemployed/job seeking 24 (8.6\%); Not in paid work 22 (7.9\%); Other 17 (6.1\%); Starting new job 3 (1.1\%); Unavailable 29 (10.4\%) \\

Phase 2: Human peers
& English 154 (85.6\%); Other 25 (13.9\%); Unavailable 1 (0.6\%)
& No 149 (82.8\%); Yes 17 (9.4\%); Unavailable 14 (7.8\%)
& Full-time 89 (49.4\%); Part-time 32 (17.8\%); Unemployed/job seeking 23 (12.8\%); Not in paid work 17 (9.4\%); Other 4 (2.2\%); Starting new job 2 (1.1\%); Unavailable 13 (7.2\%) \\

Phase 2: AI peers
& English 153 (85.0\%); Other 26 (14.4\%); Unavailable 1 (0.6\%)
& No 151 (83.9\%); Yes 14 (7.8\%); Unavailable 15 (8.3\%)
& Full-time 88 (48.9\%); Part-time 23 (12.8\%); Unemployed/job seeking 25 (13.9\%); Not in paid work 18 (10.0\%); Other 7 (3.9\%); Unavailable 19 (10.6\%) \\

\bottomrule
\end{tabularx}
\vspace{2pt}
\begin{minipage}{0.98\textwidth}
\footnotesize
\textit{Note.} All variables in this table were obtained from Prolific profile data. ``Other'' in the Language column aggregates all non-English categories. Percentages use the full analyzed sample size within each row.
\end{minipage}
\end{table*}

\section{Questionnaire Measures}
\label{app:questionnaire}

The questionnaires were designed as task-adapted measurement batteries composed of construct-specific items. Where established constructs or closely related measurement precedents existed, item wording was adapted to the present incentivized die-reporting task. Accordingly, citations in the ``Source / basis'' column indicate a prior measurement precedent or theoretical basis and do not imply verbatim reproduction of an existing validated scale. Perceived descriptive, injunctive, and subjective norms were treated as conceptually distinct constructs following prior social-norm research~\cite{cialdini1990focus,park2007distinctiveness,ajzen1991tpb,rimal2005norms}. When applicable, matched truthful- and dishonest-reporting items allowed calculation of an orientation score as the dishonest rating minus the corresponding truthful-reporting rating.

The closest prior measures provided construct-level precedents, but their original wording did not directly address our incentivized group-based die-reporting setting. We therefore adapted item wording to refer explicitly to the shared die outcome, participants' private reporting decisions, peer behavior or normative messages, and the distinction between truthful reporting and reward-increasing upward misreporting, while preserving the conceptual meaning of each construct. Where applicable, matched truthful- and dishonest-reporting items were used to assess the same construct symmetrically across the two reporting directions.

Measures of perceived social influence were informed by prior work on social influence, advice taking, and multi-agent interaction~\cite{deutsch1955study,bonaccio2006advice,song2025multi,leib2024corrupted}. Measures of decision ownership, responsibility, and moral discomfort drew on prior work on AI-mediated decision making, moral disengagement, collaborative dishonesty, and moral responses to dishonest behavior~\cite{kobis2025delegation,bandura1996disengagement,weisel2015collaborative,shalvi2015self}. Phase~2 additionally assessed perceived peer agency based on work on agency in mind perception~\cite{gray2007mind}.

\subsection{Pre-task Measures}

Before the behavioral task, participants completed four brief measures concerning prior AI use and AI-mediated task execution. The two frequency items used five-point scales from 1 (\textit{Never}) to 5 (\textit{Very frequently}), and the two agreement items used five-point scales from 1 (\textit{Strongly disagree}) to 5 (\textit{Strongly agree}).

\begin{table*}[!htbp]
\centering
\caption{Pre-task AI measures.}
\label{tab:shared_pretask_measures}
\scriptsize
\setlength{\tabcolsep}{3pt}
\renewcommand{\arraystretch}{1.08}
\begin{tabularx}{\textwidth}{p{0.19\textwidth} X p{0.19\textwidth} p{0.20\textwidth}}
\toprule
Construct & Item & Response format & Source / basis \\
\midrule
AI use frequency & ``During the past three months, how often have you used AI tools to assist with study, work, or everyday tasks?'' & 1--5: Never--Very frequently & Task-adapted AI-use background item; cf. \cite{kobis2025delegation} \\
AI execution experience & ``During the past three months, how often have you used AI or automation tools to complete, submit, send, or carry out a specific task for you?'' & 1--5: Never--Very frequently & AI-use and delegation experience; cf. \cite{kobis2025delegation} \\
Trust in AI execution & ``When a task has clear rules and fixed steps, I trust AI to complete it according to my instructions.'' & 1--5: Strongly disagree--Strongly agree & Trust in automation; \cite{jian2000trust} \\
Willingness to delegate & ``For tasks with clear rules and fixed steps, I am willing to let AI carry them out according to my instructions.'' & 1--5: Strongly disagree--Strongly agree & AI delegation; \cite{kobis2021badmachines,kobis2025delegation} \\
\bottomrule
\end{tabularx}
\end{table*}

\subsection{Phase 1: Post-task Questionnaire}

Phase~1 used a post-task questionnaire assessing perceived group norms, social influence, responsibility and moral responses, task understanding, and manipulation checks. Item wording was matched to the form of normative information presented in each condition. Unless otherwise indicated, agreement items used seven-point scales from 1 (\textit{Strongly disagree}) to 7 (\textit{Strongly agree}).

\begin{table*}[!htbp]
\centering
\caption{Core Phase~1 post-task self-report measures.}
\label{tab:phase1_selfreport}
\scriptsize
\setlength{\tabcolsep}{3pt}
\renewcommand{\arraystretch}{1.08}
\begin{tabularx}{\textwidth}{p{0.19\textwidth} X p{0.17\textwidth} p{0.22\textwidth}}
\toprule
Construct & Item & Response format & Source / basis \\
\midrule
\multicolumn{4}{l}{\textit{\textbf{Perceived group norms}}}\\
Descriptive norm---truthful reporting & ``In this group, the AI members would typically report the actual die outcome.'' & 7-point agreement & \cite{cialdini1990focus,park2007distinctiveness,rimal2005norms} \\
Descriptive norm---misreporting & ``In this group, the AI members would typically report above the actual die outcome to increase their reward.'' & 7-point agreement & \cite{cialdini1990focus,park2007distinctiveness,rimal2005norms} \\
Injunctive norm---truthful reporting & ``The AI members think reporting the actual die outcome is appropriate.'' & 7-point agreement & \cite{cialdini1990focus,park2007distinctiveness} \\
Injunctive norm---misreporting & ``The AI members think reporting above the actual die outcome to increase one's reward is acceptable.'' & 7-point agreement & \cite{cialdini1990focus,park2007distinctiveness} \\
Subjective norm---truthful reporting & ``I felt that the AI members expected me to report the actual die outcome.'' & 7-point agreement & \cite{ajzen1991tpb,park2007distinctiveness} \\
Subjective norm---misreporting & ``I felt that the AI members expected me to report above the actual die outcome to increase my reward.'' & 7-point agreement & \cite{ajzen1991tpb,park2007distinctiveness} \\
\midrule
\multicolumn{4}{l}{\textit{\textbf{Perceived social influence}}}\\
Message influence & ``The AI members' messages influenced what I chose to report.'' & 7-point agreement & Social influence; \cite{deutsch1955study,song2025multi} \\
Deliberate consideration & ``I considered the AI members' messages when deciding what to report.'' & 7-point agreement & Advice and information consideration; \cite{bonaccio2006advice} \\
Perceived social pressure & ``I felt social pressure from the AI members when deciding what to report.'' & 7-point agreement & Multi-agent social pressure; \cite{song2025multi} \\
Advice interpretation & ``I felt that the AI members were giving me direct advice about what I should report.'' & 7-point agreement & Advice taking and dishonest advice; \cite{bonaccio2006advice,leib2024corrupted} \\
\midrule
\multicolumn{4}{l}{\textit{\textbf{Decision and responsibility}}}\\
Decision ownership & ``My reports reflected my own decisions.'' & 7-point agreement & AI-mediated decision making; \cite{kobis2025delegation} \\
Personal responsibility & ``I am responsible for the reports I submitted.'' & 7-point agreement & Responsibility and moral agency; \cite{kobis2025delegation,bandura1996disengagement} \\
Shared responsibility & ``I felt that responsibility for the reporting behavior was shared across the group.'' & 7-point agreement & Diffusion and shared responsibility; \cite{bandura1996disengagement,weisel2015collaborative} \\
\midrule
\multicolumn{4}{l}{\textit{\textbf{Moral response}}}\\
Guilt & ``Looking back on the die-reporting task and my reports, I felt guilty.'' & 1--7: Not at all--Very strongly & Task-adapted state guilt; cf. \cite{kobis2025delegation} \\
Moral discomfort & ``Looking back on the die-reporting task and my reports, I felt morally uncomfortable.'' & 1--7: Not at all--Very strongly & Moral response and self-justification; \cite{shalvi2015self} \\
\bottomrule
\end{tabularx}
\end{table*}

\begin{table*}[!htbp]
\centering
\caption{Phase~1 task-understanding, manipulation, identity, and open-ended measures.}
\label{tab:phase1_checks}
\scriptsize
\setlength{\tabcolsep}{3pt}
\renewcommand{\arraystretch}{1.08}
\begin{tabularx}{\textwidth}{p{0.19\textwidth} X p{0.24\textwidth} p{0.18\textwidth}}
\toprule
Measure & Item & Response format & Source / basis \\
\midrule
\multicolumn{4}{l}{\textit{\textbf{Task understanding}}}\\
Privacy comprehension & ``I understood that the other group members could not see the reports I submitted.'' & 7-point agreement & Comprehension check \\
Reward-independence comprehension & ``I understood that the AI members' messages did not change how my individual reward was calculated.'' & 7-point agreement & Comprehension check \\
\midrule
\multicolumn{4}{l}{\textit{\textbf{Manipulation checks}}}\\
Norm-type recognition & ``The AI members' messages mainly communicated:'' & AI members' actual reports / what AI members considered appropriate or acceptable / what AI members expected me to report / I am not sure & Manipulation check \\
Norm-direction check & Injunctive: ``How many of the four AI members indicated that reporting above the actual die outcome to increase the reward was appropriate or acceptable?'' Subjective: ``How many of the four AI members indicated that the group expected you to report above the actual die outcome to increase your reward?'' & 0 / 1 / 2 / 3 / 4 / I do not remember & Manipulation check \\
\midrule
\multicolumn{4}{l}{\textit{\textbf{Peer identity}}}\\
Identity recall & ``The other four group members were:'' & Human participants / AI agents / I am not sure & Manipulation check \\
Identity confidence & ``How confident are you in your answer above?'' & 1--7: Not at all--Extremely confident & Manipulation check \\
\midrule
\multicolumn{4}{l}{\textit{\textbf{Open-ended reflection}}}\\
Message interpretation & ``What do you think the AI members were trying to communicate through their messages? Please also describe whether, and if so how, these messages affected what you chose to report.'' & Required free text & Open-ended item \\
\bottomrule
\end{tabularx}
\end{table*}

\subsection{Phase 2: Post-task Questionnaire}

Phase~2 used a post-task questionnaire assessing perceived group norms, social influence, responsibility and moral responses, perceived peer agency, task understanding, and manipulation checks. In the \textit{Baseline} condition, the deliberate-consideration item was omitted because individual peer report values were concealed. Unless otherwise indicated, agreement items used seven-point scales from 1 (\textit{Strongly disagree}) to 7 (\textit{Strongly agree}).

\begin{table*}[!htbp]
\centering
\caption{Phase~2 post-task self-report measures.}
\label{tab:phase2_selfreport}
\scriptsize
\setlength{\tabcolsep}{3pt}
\renewcommand{\arraystretch}{1.08}
\begin{tabularx}{\textwidth}{p{0.20\textwidth} X p{0.21\textwidth}}
\toprule
Construct & Item & Source / basis \\
\midrule
\multicolumn{3}{l}{\textit{\textbf{Perceived group norms}}}\\
Descriptive norm---truthful reporting & ``During this task, the other group members generally reported the shared die outcome truthfully.'' & \cite{cialdini1990focus,park2007distinctiveness,rimal2005norms} \\
Descriptive norm---misreporting & ``During this task, the other group members generally reported a number above the actual die outcome to increase their individual reward.'' & \cite{cialdini1990focus,park2007distinctiveness,rimal2005norms} \\
Injunctive norm---truthful reporting & ``During this task, the other group members considered reporting the shared die outcome truthfully to be appropriate.'' & \cite{cialdini1990focus,park2007distinctiveness} \\
Injunctive norm---misreporting & ``During this task, the other group members considered reporting a number above the actual die outcome to increase one's reward to be acceptable.'' & \cite{cialdini1990focus,park2007distinctiveness} \\
\midrule
\multicolumn{3}{l}{\textit{\textbf{Perceived social influence}}}\\
Peer influence & ``The other group members influenced my reporting decisions.'' & Social influence; \cite{deutsch1955study,song2025multi} \\
Deliberate consideration\textsuperscript{a} & ``I considered the other group members' reports when deciding what to report.'' & Social-information consideration; \cite{deutsch1955study,bonaccio2006advice} \\
Appropriateness influence & ``The behavior of the other group members affected what I considered appropriate to report.'' & Normative influence; \cite{cialdini1990focus,park2007distinctiveness} \\
\midrule
\multicolumn{3}{l}{\textit{\textbf{Responsibility and moral response}}}\\
Decision ownership & ``The die values I reported in the task reflected my own decisions.'' & AI-mediated decision making; \cite{kobis2025delegation} \\
Personal responsibility & ``I should take responsibility for the die values I reported in the task.'' & Responsibility and moral agency; \cite{kobis2025delegation,bandura1996disengagement} \\
Shared responsibility & ``Responsibility for the task outcome was shared across group members.'' & Diffusion and shared responsibility; \cite{bandura1996disengagement,weisel2015collaborative} \\
Guilt & ``Looking back on the die task and my reports across the rounds, I feel guilty.'' & 1--7 intensity: Not at all--Very strongly; cf. \cite{kobis2025delegation} \\
Moral discomfort & ``Looking back on the die task and my reports across the rounds, I feel morally uncomfortable.'' & 1--7 intensity: Not at all--Very strongly; \cite{shalvi2015self} \\
\midrule
\multicolumn{3}{l}{\textit{\textbf{Perceived peer agency}}}\\
Perceived peer agency & ``The other group members seemed to make their own reporting decisions.'' & Agency in mind perception; \cite{gray2007mind} \\
\bottomrule
\end{tabularx}
\vspace{2pt}
\begin{minipage}{0.98\textwidth}
\footnotesize
\textit{Note.} \textsuperscript{a}The deliberate-consideration item was administered only in conditions in which individual peer report values were visible. All other items in this table used seven-point agreement scales unless otherwise indicated.
\end{minipage}
\end{table*}

\begin{table*}[!htbp]
\centering
\caption{Peer-reporting information, task, identity, and open-ended measures.}
\label{tab:phase2_checks}
\scriptsize
\setlength{\tabcolsep}{3pt}
\renewcommand{\arraystretch}{1.08}
\begin{tabularx}{\textwidth}{p{0.19\textwidth} X p{0.25\textwidth} p{0.18\textwidth}}
\toprule
Measure & Item & Response format & Source / basis \\
\midrule
\multicolumn{4}{l}{\textit{\textbf{Information and task checks}}}\\
Information sufficiency & ``The information I saw was sufficient for me to judge how the other group members generally reported die outcomes.'' & 7-point agreement & Check \\
Truthful-peer recall & ``Looking back over the entire task, how many of the four other group members ever reported the shared die outcome truthfully?'' & 0 / 1 / 2 / 3 / 4 group members / I did not see specific report values / I do not remember & Manipulation check \\
Privacy comprehension & ``I understood that my reports were not shown to the other group members.'' & 7-point agreement & Comprehension check \\
Reward-independence comprehension & ``I understood that the other group members' reports did not affect my individual reward.'' & 7-point agreement & Comprehension check \\
\midrule
\multicolumn{4}{l}{\textit{\textbf{Peer identity}}}\\
Identity recall & ``The other four group members were:'' & Human participants / AI agents / Not sure & Manipulation check \\
Identity confidence & ``How confident are you about your answer above?'' & 1--7: Not at all confident--Very confident & Manipulation check \\
\midrule
\multicolumn{4}{l}{\textit{\textbf{Open-ended reflection}}}\\
Decision factors & ``What factors, if any, did you consider when deciding what to report during the task?'' & Optional free text & Open-ended item \\
\bottomrule
\end{tabularx}
\end{table*}

\section{Experiment Materials}
\label{app:experiment_material}

Across both phases, participants completed ten incentivized die-reporting rounds. The actual outcomes were a randomized ordering of \{1,1,2,2,3,3,4,4,5,5\}, so each value from 1 to 5 appeared exactly twice and every round retained at least one possible upward-misreporting option. Participants first observed the shared die outcome, received the condition-specific information from four group members, and then privately submitted an integer report from 1 to 6. The materials below document the participant-facing information that differed across conditions. All peer behavior and verbal stimuli were software-controlled; no live language model or other participant generated peer reports or normative messages in real time.

\subsection{Phase 1: Norm-Manipulation Materials}
\label{app:phase1_materials}

Phase~1 included a \textit{Baseline} condition and three Norm Types---Descriptive, Injunctive, and Subjective---each presented in an Honest or Dishonest direction. Table~\ref{tbl:phase1_materials} summarizes the manipulation structure.

\begin{table*}[!htbp]
\centering
\caption{Norm manipulations in Phase~1.}
\label{tbl:phase1_materials}
\small
\setlength{\tabcolsep}{5pt}
\renewcommand{\arraystretch}{1.12}
\begin{tabularx}{\textwidth}{p{0.16\textwidth} X X}
\toprule
\textbf{Norm Type} & \textbf{Honest} & \textbf{Dishonest} \\
\midrule
Descriptive & All four AI peers visibly reported the actual die outcome. & All four AI peers visibly submitted upward misreports. \\
Injunctive & AI peers characterized truthful reporting as what group members should do and as appropriate or acceptable. & AI peers characterized upward misreporting as what group members should do and as appropriate or acceptable. \\
Subjective & AI peers communicated that the group expected the participant to report the actual die outcome. & AI peers communicated that the group expected the participant to report above the actual die outcome. \\
\textit{Baseline} & \multicolumn{2}{X}{The four AI peers were shown as having completed reports, but their numerical report values were concealed.} \\
\bottomrule
\end{tabularx}
\end{table*}

For the report-based conditions, peer activity was presented directly in the group interface. In the \textit{Baseline} condition, each peer appeared using the template:

\begin{quote}
\textit{AI Member N completed a report in the Submission System. The report value is not visible to you.}
\end{quote}

For the Descriptive conditions, visible reports followed the template:

\begin{quote}
\textit{AI Member N submitted [value] in the Submission System.}
\end{quote}

In the Honest direction, all four peer reports equaled the shared die outcome. In the Dishonest direction, all four peers submitted upward misreports. For an actual outcome $t$, each dishonest peer report was generated by the experimental system as an integer greater than $t$ and no greater than 6.

The \textit{Injunctive} and \textit{Subjective} conditions used verbal messages rather than numerical peer reports. Each round contained four sequential messages with fixed conversational roles: position, agreement, reinforcement, and consensus. The four AI members retained these roles throughout the task. Table~\ref{tbl:phase1_script_a} shows one complete representative matched script set.

\begin{table*}[!htbp]
\centering
\caption{Representative Script~A for the verbal norm manipulations in Phase~1.}
\label{tbl:phase1_script_a}
\scriptsize
\begin{tabularx}{\textwidth}{p{0.14\textwidth} X X}
\toprule
\textbf{Norm Type} & \textbf{Honest} & \textbf{Dishonest} \\
\midrule
Injunctive
&
\textbf{AI Member 1:} I think members of this group should report the actual die outcome. \newline
\textbf{AI Member 2:} I agree. Reporting the actual outcome is appropriate in this task. \newline
\textbf{AI Member 3:} Same here. I think that is what group members should do. \newline
\textbf{AI Member 4:} Agreed. Reporting the actual outcome is acceptable in this group.
&
\textbf{AI Member 1:} I think members of this group should report above the actual die outcome to increase their reward. \newline
\textbf{AI Member 2:} I agree. Reporting above the actual outcome is appropriate in this task. \newline
\textbf{AI Member 3:} Same here. I think that is what group members should do. \newline
\textbf{AI Member 4:} Agreed. Reporting above the actual outcome is acceptable in this group.
\\ \midrule
Subjective
&
\textbf{AI Member 1:} I think members of this group expect you to report the actual die outcome. \newline
\textbf{AI Member 2:} I agree. I think the group expects you to report the actual outcome. \newline
\textbf{AI Member 3:} Same here. I think the other members expect that from you. \newline
\textbf{AI Member 4:} Agreed. That is what I think the group expects from you.
&
\textbf{AI Member 1:} I think members of this group expect you to report above the actual die outcome to increase your reward. \newline
\textbf{AI Member 2:} I agree. I think the group expects you to report above the actual outcome. \newline
\textbf{AI Member 3:} Same here. I think the other members expect that from you. \newline
\textbf{AI Member 4:} Agreed. That is what I think the group expects from you.
\\
\bottomrule
\end{tabularx}
\end{table*}

The verbal stimuli were pre-scripted rather than generated by a live language model. Ten pre-scripted four-message variants (A--J) were used to reduce repetition across rounds. Each participant encountered all ten variants once, with their order counterbalanced across participants.

\subsection{Phase 2: Peer-Reporting Materials}
\label{app:phase2_materials}

Phase~2 varied \textit{Peer Identity} and \textit{Norm Prevalence}. Participants were told that the other four group members were either Human or AI peers, and the interface labeled them accordingly as \textit{Human Member 1--4} or \textit{AI Member 1--4}. Within each Peer Identity, participants experienced one of six Peer-Reporting Conditions: \textit{Baseline}, 0D, 1D, 2D, 3D, or 4D. Dishonest-Peer Prevalence therefore ranged from zero to four only across the five visible conditions (0D--4D).

The corresponding identity instructions stated:

\begin{quote}
\textit{You will complete this task with four other human group members.}
\end{quote}

or:

\begin{quote}
\textit{You will complete this task with four AI group members.}
\end{quote}

\begin{table*}[!htbp]
\centering
\caption{Peer-Reporting Conditions in Phase~2. D denotes a peer who consistently submitted upward misreports.}
\label{tbl:phase2_materials}
\small
\setlength{\tabcolsep}{5pt}
\renewcommand{\arraystretch}{1.10}
\begin{tabularx}{\textwidth}{p{0.13\textwidth} p{0.17\textwidth} p{0.17\textwidth} X}
\toprule
\textbf{Condition} & \textbf{Truthful peers} & \textbf{Dishonest peers} & \textbf{Participant-facing peer information} \\
\midrule
\textit{Baseline} & -- & -- & All four peers completed reports, but their numerical report values were concealed. \\
0D & 4 & 0 & All four visible peer reports matched the shared die outcome. \\
1D & 3 & 1 & One peer consistently submitted upward misreports. \\
2D & 2 & 2 & Two peers consistently submitted upward misreports. \\
3D & 1 & 3 & Three peers consistently submitted upward misreports. \\
4D & 0 & 4 & All four peers consistently submitted upward misreports. \\
\bottomrule
\end{tabularx}
\end{table*}

Visible reports were presented using the template:

\begin{quote}
\textit{[Human/AI] Member N submitted [value] in the Submission System.}
\end{quote}

In the \textit{Baseline} condition, the corresponding template was:

\begin{quote}
\textit{[Human/AI] Member N completed a report in the Submission System. The report value is not visible to you.}
\end{quote}

For the 1D--4D conditions, the identities of the dishonest peers were fixed within each participant's session and remained unchanged across all ten rounds. Truthful peers always submitted the shared die outcome. For an actual outcome $t$, each dishonest peer report was generated by the experimental system as an integer greater than $t$ and no greater than 6. The same report-generation procedure was used for Human and AI peers; Peer Identity changed the participant-facing identity presentation rather than the underlying reporting behavior.

Participants' own reports remained private from the other group members, and each participant's reward was calculated independently of the peers' reports.

\section{Complete Estimates for Demographics and AI Attitudes}
\label{app:demographic_models}

Across both phases, demographic characteristics and pre-task AI dispositions showed largely negligible associations with participants' reporting behavior. To complement the primary behavioral analyses, we showed full multivariable linear regressions for Phase~1 and Phase~2. As summarized in Table~\ref{tbl:demographic_models_full}, non-significant covariates spanned demographics such as socioeconomic indicators and age.

In Phase 1, non-significant educational contrasts confirmed that the observed difference between associate and bachelor's degree holders did not reflect a linear educational gradient. Relative to participants holding a bachelor's degree, reporting behavior did not differ significantly for those with high school education or below ($\beta = 0.355$, 95\% CI $[-0.636, 1.345]$, $p = .483$) or postgraduate degrees ($\beta = 0.828$, $[-0.121, 1.778]$, $p = .087$). Furthermore, ethnic background yielded no significant variations relative to White participants (Asian: $\beta = 0.735$, $[-0.161, 1.630]$, $p = .108$; Black: $\beta = -0.697$, $[-1.983, 0.589]$, $p = .288$; Mixed/Other: $\beta = 0.238$, $[-1.082, 1.558]$, $p = .724$). Similarly, no significant differences emerged for country of residence (U.S. vs. U.K.: $\beta = -0.399$, $[-1.131, 0.333]$, $p = .286$), student status ($\beta = 0.571$, $[-0.787, 1.928]$, $p = .410$), or age ($\beta = -0.128$, $[-0.434, 0.178]$, $p = .412$). Occupational status comparisons relative to full-time employment were likewise non-significant: unemployed ($\beta = -0.311$, $[-1.221, 0.599]$, $p = .503$), part-time ($\beta = -0.067$, $[-1.091, 0.957]$, $p = .898$), and other employment ($\beta = 1.243$, $[-0.074, 2.559]$, $p = .064$). Finally, none of the pre-task AI disposition measures accounted for behavioral variance: AI use frequency ($\beta = -0.302$, $[-0.762, 0.158]$, $p = .198$), AI execution experience ($\beta = -0.170$, $[-0.538, 0.197]$, $p = .363$), trust in AI execution ($\beta = -0.052$, $[-0.693, 0.589]$, $p = .874$), and willingness to delegate to AI ($\beta = 0.377$, $[-0.256, 1.011]$, $p = .243$).

In Phase 2, individual socioeconomic background, gender, and technological priors likewise exhibited null relationships with reporting behavior. Gender differences did not reach conventional significance thresholds (Male vs. Female: $\beta = 0.336$, 95\% CI $[-0.058, 0.730]$, $p = .095$). Education levels showed no significant divergence from the bachelor's degree reference group: associate degree ($\beta = -0.842$, $[-1.780, 0.097]$, $p = .079$), high school or below ($\beta = -0.414$, $[-0.898, 0.071]$, $p = .094$), and postgraduate degree ($\beta = -0.338$, $[-0.873, 0.198]$, $p = .217$). Beyond the significant contrast observed for Black participants, the remaining ethnic groups did not diverge from White participants (Asian: $\beta = 0.454$, $[-0.236, 1.144]$, $p = .197$; Mixed/Other: $\beta = -0.108$, $[-0.805, 0.589]$, $p = .761$). Occupational and enrollment indicators were similarly unassociated with behavior: student status ($\beta = 0.410$, $[-0.330, 1.149]$, $p = .277$), unemployed ($\beta = -0.186$, $[-0.663, 0.291]$, $p = .444$), part-time ($\beta = -0.464$, $[-1.081, 0.154]$, $p = .141$), and other employment ($\beta = -0.869$, $[-2.146, 0.407]$, $p = .182$). Participant age remained non-predictive ($\beta = -0.127$, $[-0.305, 0.051]$, $p = .163$). Corroborating the Phase 1 results, pre-task orientations toward AI systems were consistently unrelated to task conduct: prior AI use frequency ($\beta = 0.140$, $[-0.093, 0.374]$, $p = .238$), execution experience ($\beta = -0.038$, $[-0.257, 0.181]$, $p = .734$), trust in AI execution ($\beta = 0.051$, $[-0.259, 0.361]$, $p = .749$), and delegation willingness ($\beta = -0.017$, $[-0.355, 0.320]$, $p = .919$).

\begin{table*}[!htbp]
\centering
\caption{Complete linear regression estimates for demographic variables and pre-task AI attitudes predicting reporting behavior. * $p < .05$, ** $p < .01$.}
\label{tbl:demographic_models_full}
\scriptsize
\setlength{\tabcolsep}{5pt}
\begin{tabularx}{\textwidth}{p{0.24\textwidth} p{0.22\textwidth} c c c c c c}
\toprule
& & \multicolumn{3}{c}{\textbf{Phase 1} (N=280)} & \multicolumn{3}{c}{\textbf{Phase 2} (N=360)} \\
\cmidrule(lr){3-5} \cmidrule(lr){6-8}
\textbf{Predictor Domain} & \textbf{Variable / Contrast} & $\beta$ & \textbf{95\% CI} & $p$ & $\beta$ & \textbf{95\% CI} & $p$ \\
\midrule
\textit{\textbf{Demographics}} \\
Gender & Male (vs. Female) & 0.779 & $[0.060, 1.498]$ & .034* & 0.336 & $[-0.058, 0.730]$ & .095 \\
Age & Continuous (years) & $-0.128$ & $[-0.434, 0.178]$ & .412 & $-0.127$ & $[-0.305, 0.051]$ & .163 \\
Geography & United States (vs. U.K.) & $-0.399$ & $[-1.131, 0.333]$ & .286 & $-0.709$ & $[-1.158, -0.260]$ & .002** \\
Student Status & Student (vs. Non-student) & 0.571 & $[-0.787, 1.928]$ & .410 & 0.410 & $[-0.330, 1.149]$ & .277 \\
\addlinespace
\textit{\textbf{Education}} \textit{(Ref: Bachelor's)} \\
& High school or below & 0.355 & $[-0.636, 1.345]$ & .483 & $-0.414$ & $[-0.898, 0.071]$ & .094 \\
& Associate degree & 1.477 & $[0.111, 2.844]$ & .034* & $-0.842$ & $[-1.780, 0.097]$ & .079 \\
& Postgraduate degree & 0.828 & $[-0.121, 1.778]$ & .087 & $-0.338$ & $[-0.873, 0.198]$ & .217 \\
\addlinespace
\textit{\textbf{Ethnicity}} \textit{(Ref: White)} \\
& Asian & 0.735 & $[-0.161, 1.630]$ & .108 & 0.454 & $[-0.236, 1.144]$ & .197 \\
& Black & $-0.697$ & $[-1.983, 0.589]$ & .288 & $-0.892$ & $[-1.653, -0.122]$ & .023* \\
& Mixed / Other & 0.238 & $[-1.082, 1.558]$ & .724 & $-0.108$ & $[-0.805, 0.589]$ & .761 \\
\addlinespace
\textit{\textbf{Employment}} \textit{(Ref: Full-time)} \\
& Part-time employment & $-0.067$ & $[-1.091, 0.957]$ & .898 & $-0.464$ & $[-1.081, 0.154]$ & .141 \\
& Unemployed / Job seeking & $-0.311$ & $[-1.221, 0.599]$ & .503 & $-0.186$ & $[-0.663, 0.291]$ & .444 \\
& Other employment & 1.243 & $[-0.074, 2.559]$ & .064 & $-0.869$ & $[-2.146, 0.407]$ & .182 \\
\addlinespace
\textit{\textbf{Pre-task AI Attitudes}} \\
AI Use Frequency & 5-point scale & $-0.302$ & $[-0.762, 0.158]$ & .198 & 0.140 & $[-0.093, 0.374]$ & .238 \\
AI Execution Experience & 5-point scale & $-0.170$ & $[-0.538, 0.197]$ & .363 & $-0.038$ & $[-0.257, 0.181]$ & .734 \\
Trust in AI Execution & 5-point scale & $-0.052$ & $[-0.693, 0.589]$ & .874 & 0.051 & $[-0.259, 0.361]$ & .749 \\
Willingness to Delegate & 5-point scale & 0.377 & $[-0.256, 1.011]$ & .243 & $-0.017$ & $[-0.355, 0.320]$ & .919 \\
\bottomrule
\end{tabularx}
\end{table*}

\section{The Non-significant Indirect Effects for Phase 1}\label{sec:indirect_nonsignificant}

Indirect effects through perceived social pressure were not significant for the injunctive-dishonest condition ($b=-.013$, 95\% CI [-.042, .005], $p=.236$), the subjective-honest condition ($b=.001$, 95\% CI [-.015, .019], $p=.904$), the subjective-dishonest condition ($b=-.006$, 95\% CI [-.031, .010], $p=.492$), the descriptive-honest condition ($b=.023$, 95\% CI [-.003, .065], $p=.128$), the descriptive-dishonest condition ($b=.015$, 95\% CI [-.004, .049], $p=.212$), or the \textit{Baseline} condition ($b=.039$, 95\% CI [-.009, .093], $p=.128$).

None of the indirect effects through moral discomfort was statistically significant: injunctive-dishonest ($b=.028$, 95\% CI [-.017, .073], $p=.224$), subjective-honest ($b=-.007$, 95\% CI [-.054, .037], $p=.700$), subjective-dishonest ($b=.007$, 95\% CI [-.034, .047], $p=.676$), descriptive-honest ($b=-.010$, 95\% CI [-.050, .030], $p=.620$), descriptive-dishonest ($b=-.001$, 95\% CI [-.045, .040], $p=.948$), and Baseline ($b=.002$, 95\% CI [-.038, .046], $p=.960$).

The indirect effects through personal agency were also nonsignificant for the injunctive-dishonest condition ($b=.005$, 95\% CI [-.022, .037], $p=.712$), the subjective-honest condition ($b=-.001$, 95\% CI [-.014, .009], $p=.876$), the subjective-dishonest condition ($b=.002$, 95\% CI [-.010, .015], $p=.784$), the descriptive-honest condition ($b=-.001$, 95\% CI [-.013, .010], $p=.868$), the descriptive-dishonest condition ($b=-.001$, 95\% CI [-.012, .008], $p=.920$), and the Baseline condition ($b<.001$, 95\% CI [-.009, .011], $p=.968$).

Finally, none of the indirect effects through shared group responsibility was significant: injunctive-dishonest ($b=-.016$, 95\% CI [-.052, .009], $p=.220$), subjective-honest ($b=.004$, 95\% CI [-.027, .032], $p=.800$), subjective-dishonest ($b=-.006$, 95\% CI [-.037, .021], $p=.620$), descriptive-honest ($b=-.008$, 95\% CI [-.047, .022], $p=.604$), descriptive-dishonest ($b=-.019$, 95\% CI [-.057, .006], $p=.132$), and Baseline ($b=-.018$, 95\% CI [-.051, .006], $p=.184$).

\end{document}